\documentclass[prb, twocolumn, showpacs, amsmath, amsfonts, amssymb,superscriptaddress]{revtex4-1}
\usepackage{mysetup}
\usepackage{xstring}
\usepackage{mathtools}

\usepackage{color}

\usepackage{cancel}

\newcommand{\ci}{\textnormal{C}}
\newcommand{\li}{\textnormal{L}}
\newcommand{\ri}{\textnormal{R}}
\newcommand{\lli}{\tilde{\textnormal{L}}}
\newcommand{\rri}{\tilde{\textnormal{R}}}
\newcommand{\si}{\textnormal{S}}
\newcommand{\tn}{\textnormal}

\newcommand{\calrL}{{\cal G}^{\textnormal{ret},\cancel{\li}}}
\newcommand{\calrR}{{\cal G}^{\textnormal{ret},\cancel{\ri}}}
\newcommand{\calrLR}{{\cal G}^{\textnormal{ret},\cancel{\li}\cancel{\ri}}}

\newcommand{\calaL}{{\cal G}^{\textnormal{adv},\cancel{\li}}}
\newcommand{\calaR}{{\cal G}^{\textnormal{adv},\cancel{\ri}}}
\newcommand{\calaLR}{{\cal G}^{\textnormal{adv},\cancel{\li}\cancel{\ri}}}

\newcommand{\calkLR}{{\cal G}^{\textnormal{K},\cancel{\li}\cancel{\ri}}}

\newcommand{\tsk}{\tilde\Sigma^{\tn{K}}}

\newcommand{\calSrR}{{\cal S}^{\textnormal{ret},\cancel{\ri}}}
\newcommand{\calSrLR}{{\cal S}^{\textnormal{ret},\cancel{\li}\cancel{\ri}}}

\newcommand{\calSaR}{{\cal S}^{\textnormal{adv},\cancel{\ri}}}
\newcommand{\calSaLR}{{\cal S}^{\textnormal{adv},\cancel{\li}\cancel{\ri}}}

\newcommand{\calSkLR}{{\cal S}^{\textnormal{K},\cancel{\li}\cancel{\ri}}}

\newcommand{\pa}{\partial_\Lambda^*}

\newcommand{\syst}[2]{
\begin{tikzpicture}
    \def\efi{#1}
    \coordinate(A0) at (0.6cm,0cm);
    \foreach \a in {1,2,...,7}{
        \if\a1%
            \node[circle, minimum size=0.6cm, below right=\efi*\a and 1.2cm*\a of A0, thick](A\a){$\dots$};
        \else
            \if\a7%
                \node[circle, minimum size=0.6cm, below right=\efi*\a and 1.2cm*\a of A0, thick](A\a){$\dots$};
            \else
                \node[draw, circle, minimum size=0.6cm, below right= \efi*\a and 1.2cm*\a of A0, thick](A\a){};
            \fi
        \fi;
    }
    \foreach[evaluate=\a as \an using int(\a+1)] \a in {1,2,...,6}
        \path[]
            (A\a.north east) edge [
              text=black,
              thick,
              shorten <=2pt,
              shorten >=2pt,
              bend left=50
              ] node[above]{\IfEqCase{#2}{%
                              {1}{\(t\)}%
                              {0}{\(\)}%
                             }%
                           } (A\an.north west);
    \foreach[evaluate=\a as \an using int(\a+1)] \a in {1,2,...,6}
        \path[]
            (A\a.south east) edge [
              text=black,
              thick,
              shorten <=2pt,
              shorten >=2pt,
              bend right=50,
              dashed
              ] node[below]{\IfEqCase{#2}{%
                              {1}{\(U\)}%
                              {0}{\(\)}%
                             }%
                           } (A\an.south west);

    \foreach \a in {2,3,...,6}{
          \coordinate[below=2.6cm of A\a](r\a);
          \draw[thick, shading=axis, bottom color=white, top color=custBlue] (r\a)++(-0:0.4 and 1.5) arc(-0:180:0.4 and 1.5);
        \path[]
            (r\a)++(90:0.4 and 1.5) edge [
              text=black,
              thick,
              shorten <=2pt,
              shorten >=2pt,
              bend right=50
              ] node[left]{\IfEqCase{#2}{%
                              {1}{\(\Gamma\)}%
                              {0}{\(\)}%
                             }%
                           } (A\a.south);
    }
    \node[above=1.5cm of A2] (ar){};
    \draw[->] (ar) --node[above]{$E$} +(1.2cm,-\efi);
\end{tikzpicture}
}

\begin{document}
\title{Phases of translation-invariant systems out of equilibrium: Iterative Green's function techniques and renormalization group approaches}

\author{C. Kl\"ockner}
\affiliation{Technische Universit\"at Braunschweig, Institut f\"ur Mathematische Physik, Mendelssohnstraße 3, 38106 Braunschweig, Germany}

\author{D.M.\ Kennes}
\affiliation{Institut f\"ur Theorie der Statistischen Physik, RWTH Aachen University and JARA-Fundamentals of Future Information Technology, 52056 Aachen, Germany}

\affiliation{Max  Planck  Institute  for  the  Structure  and  Dynamics  of  Matter,Luruper  Chaussee  149,  22761  Hamburg,  Germany}

\author{C.\ Karrasch}
\affiliation{Technische Universit\"at Braunschweig, Institut f\"ur Mathematische Physik, Mendelssohnstraße 3, 38106 Braunschweig, Germany}

\begin{abstract} 
We introduce a method to evaluate the steady-state non-equilibrium Keldysh-Schwinger Green's functions for infinite systems subject to both an electric field and a coupling to reservoirs. The method we present exploits a physical quasi-translation invariance, where a shift by one unit cell leaves the physics invariant if all electronic energies are simultaneously shifted by the magnitude of the electric field. Our framework is straightaway applicable to diagrammatic many-body methods. We discuss two flagship applications, mean-field theories as well as a sophisticated second-order functional renormalization group approach. The latter allows us to push the renormalization-group characterization of phase transitions for lattice fermions into the out-of-equilibrium realm. We exemplify this by studying a model of spinless fermions, which in equilibrium exhibits a Berezinskii-Kosterlitz-Thouless phase transition.    
\end{abstract}

\pacs{} 
\date{\today} 
\maketitle

\section{Introduction}
Unconventional phases of matter play an integral role in condensed matter research and beyond.\cite{Sachdev2009} Understanding the conditions under which systems harboring many particles conspire to give rise to these emergent, collective phenomena is crucial from a fundamental as well as a technological perspective. The description of such phases also poses a formidable theoretical challenge as they are usually driven by  interactions and independent particle pictures fail spectacularly. To remedy this, powerful many-body techniques such as the renormalization group where developed. After years of research, much is known about the classification of phases of matter in thermal equilibrium as well as about the transitions between them.\cite{Sachdev2009} 

As a second step, one might wonder about ways of controlling these phases beyond the possibilities offered by equilibrium means.\cite{Basov2017} A particular non-equilibrium route that is routinely followed in experiments is to apply electric fields to solids. If the electric field is strong enough, the linear response regime is left, electrons are driven out of equilibrium, and non-linear effects become relevant. This so-called non-linear transport regime has attracted much interest in the last decades and many counter-intuitive effects were demonstrated. E.g., it was shown that 
a negative differential conductance\cite{Taguchi2000,Boulat2008,Inada2009,Mori2009} 
and oscillating currents (thyristor effect)\cite{Sawano2005} can arise and that this might have significant implications for highly efficient heat
engines.\cite{Mahan1997,Ojanen2008}
For very strong electric fields compared to the scattering rate of electrons, coherent Bloch oscillations are found,\cite{Bloch1928,Zener1934} which in the absence of scattering will not decay.\cite{Turkowski2005} In a metallic condensed matter setup, these oscillations are challenging to observe experimentally because the scattering-induced relaxation  is usually very fast on the time scale of the oscillation frequency and the steady-state current quickly relaxes to zero.\cite{Freericks2006,Turkowski2007,Freericks2008}  However, these oscillations can be accessed in semiconductors\cite{Glck2002} or cold-atom systems.\cite{BenDahan1996,Tarruell2012} For non-interacting electrons (i.e., in the absence of scattering), these oscillations can be understood as a gradient-field induced localization of the electron wave functions,\cite{Wannier1962} an effect known  as \textit{Wannier-Stark localization}.\cite{Aoki2014,Davison1997,Neumayer2015} Recently, it was shown that this localization might survive even when interactions are turned on,\cite{Schulz2019,vanNieuwenburg2019} yielding the concept of Stark many-body localization akin to many-body localization induced by quenched, quasi-periodic or programmable disorder.\cite{MBL1,MBL2,Schreiber17,Luschen17,Bar_Lev_2017,Enss17,Augustine2DMBL} These studies elevate closed electric-field driven quantum systems to the frontier of research concerning ergodicity breaking, and thus effects beyond the paradigm of statistical mechanics can be expected in these systems.  

However, when considering interacting closed systems (as discussed above) under an external driving force and beyond the regime of many-body localization, the drive will continuously heat up the system until an infinite temperature state is reached by the growing deposition of energy. This state is not very interesting, but fortunately a more realistic model includes infinite baths\cite{Li2015} which can dissipate this additional energy.\cite{Mierzejewski2011,Amaricci2012,Aron2012,Han2013,Han2013b} In such a setup, an interesting non-equilibrium state (supporting, e.g., a finite value of the
steady state current) is conceivable. In this context, negative differential conductance was reported if the electric field is increased at a constant coupling to the reservoirs. This negative differential conductance is a consequence of the current being suppressed for increasing fields, because the
amount of energy per unit time dissipated by the bath remains constant, and thus the effective temperature increases, which in turn decreases the current.\cite{Mierzejewski2011,Aron2012b, Amaricci2012}

Here we want to address the question of what happens to the electronic phases of matter as an increasingly strong field is driving the system out of its equilibrium state\cite{Mitra2008,Sieberer2016,Mathey2019} from a microscopic model perspective.  To this end, we extend a renormalization group approach,\cite{Metzner2012} which was successfully applied to characterize phases of matter in microscopic models in equilibrium,\cite{Markhof2018,Weidinger2019} to the non-equilibrium realm. This allows us to address how phases of matter can be controlled using non-equilibrium means via mechanism such as the  dielectric breakdown of insulators. This mechanism is only one example out of the broader class of non-equilibrium control avenues and describes that a correlation driven Mott insulating state can be turned metallic after the field strength has surpassed a certain threshold value where the metallization occurs via the production of doublon-hole pairs.\cite{Oka2003,Oka2005,HeidrichMeisner2010,Oka2010,Oka2012,Eckstein2010,Eckstein2013} While we apply the developed methodology to the case of an infinite one-dimensional nearest neighbor chain of spinless fermions, the general framework we derive allows to study non-equilibrium control of phases of matter in general tight-binding models, while keeping track of all of the microscopic details. 

The rest of this paper is structured as follows: In section \ref{sec:class_of_models}, we introduce the class of models that can be treated using our methods. After briefly recapitulating the Keldysh Green's function formalism in Sec.~\ref{sec:greens_functions}, we discuss an iterative algorithm to compute the Green's functions of an infinite system (Sec.~\ref{sec:iterative_gf}). This algorithm is widely applicable within all diagrammatic techniques such as (dynamical) mean-field theory; similar approaches in the context of Wannier-Stark localization can be found, e.g., in Refs.~\onlinecite{Davison1997,Neumayer2015}. We develop a full second-order implementation of the Keldysh functional renormalization group (that accounts for inelastic scattering) for infinite, open systems subject to an electric field in Sec.~\ref{sec:fRG}. As an example, we then apply this methodology to an interacting tight-binding chain coupled to reservoirs in an electric field (Sec.~\ref{sec:application}). We thoroughly discuss numerical details, and we investigate the survival of the charge-density wave transition when the reservoir couplings and/or an electric field are switched on.

\section{Class of models}\label{sec:class_of_models}
First, we will outline the class of systems that can be treated using our method. 
We eventually aim at modelling infinitely extended, one-dimensional chains of charged fermions which are coupled to reservoirs and which are subject to an electric field. As a starting point, we consider a general fermionic Hamiltonian 
with an infinite number of degrees of freedom that features a kinetic energy as well as a two-particle interaction:
\begin{equation}
    H_\mathrm{sys}=\sum_{i,j\in \mathbb{Z}} h_{ij}c_i^\dagger c^\vdag_j+\frac{1}{4} \sum_{i,j,k,l\in \mathbb{Z}} v_{ijkl} c_i^\dagger c_j^\dagger c^\vdag_l c^\vdag_k,
\end{equation}
where $h_{ij}=h_{ji}^*$, $v_{ijkl}=-v_{jikl}=-v_{ijlk}$, and \(c^{(\dag)}\) denote the fermionic annihilation (creation) operator. We also refer to $H_\mathrm{sys}$ as the \emph{chain}. This system is assumed to be  coupled to an infinite set of fermionic, non-interacting reservoirs:
\begin{equation}
    \begin{split}
    H^\nu_\mathrm{res}&=\sum_k \epsilon^\nu_k a_{k,\nu}^\dagger a_{k,\nu}^\vdag,\\
    H^\nu_\mathrm{coup}&=\sum_{i,k} t^\nu_{i,k} c_i^\dag a_{k,\nu}^\vdag\ +\ \mathrm{h.c.},
    \end{split}
\end{equation}
where \(a^{(\dag)}\) denote the fermionic annihilation (creation) operators within the reservoirs. The total Hamiltonian is given by:
\begin{equation}
    H_\mathrm{tot}=H_\mathrm{sys}+\sum_\nu\big[ H^\nu_\mathrm{res}+H^\nu_\mathrm{coup}\big].
\end{equation}

The initial state is assumed to be one where the reservoirs are decoupled ($H^\nu_\mathrm{coup}=0$) and are by themselves in thermal equilibrium. The influence of the reservoirs can then be characterized by the following hybridization functions (which will play the role of reservoir self-energies in the Dyson equation):
\begin{equation}\label{eq:self_hybrid}\begin{split}
    \Gamma^{\nu,\mathrm{ret}}_{ij}(\omega)&= \sum_k t^\nu_{i,k}t^{\nu*}_{j,k} \frac{1}{\omega-\epsilon^\nu_k+\I 0^+},\\
    \Gamma^{\nu,\mathrm{K}}_{ij}(\omega)&= [1-2n^\nu(\omega)]\, 2\I\, \textnormal{Im }\Gamma^{\nu,\mathrm{ret}}_{ij}(\omega) ,    
\end{split}\end{equation}
where $n^\nu(\omega)$ is a thermal (Fermi) distribution function:
\begin{equation}
n^\nu(\omega)=\frac{1}{\exp[(\omega-\mu_\nu)/T]+1}. 
\end{equation}
We also assume that all degrees of freedom within the chain feature some decay channel into the reservoirs, guaranteeing a well-defined stationary state that is independent of the initial preparation of the chain itself.

Guided by the picture of a chain in an electric field, we restrict ourselves to Hamiltonians which have a discrete translational shift symmetry.
With a given \(L\in\mathbb{N}\) defining a unit cell, we demand that
\begin{equation}
    \label{eq:shift_sym}
    \begin{split}
        h_{(i+L)(j+L)}&=h_{i j}+LE\delta_{i,j},\\
        v_{(i+L)(j+L)(k+L)(l+L)}&=v_{ijkl}\hspace{0.8cm} \forall i,j,k,l\in\mathbb{Z}.
    \end{split}
\end{equation}
The quantity \(E>0\) has the interpretation of an electric field in arbitrary units.
The hybridization and distribution function are similarly required to fulfill\footnote{It would be interesting to generalize our method to the case that the baths and the chain feature different translational shift symmetries. However, this is not straightforward.}
\begin{equation}
    \label{eq:shift_sym_res}
    \begin{split}
    \Gamma^{\nu+L,\mathrm{ret}}_{(i+L)(j+L)}(\omega)&=\Gamma^{\nu,\mathrm{ret}}_{ij}(\omega-LE)~~~~\forall i,j\in\mathbb{Z},\\
    n^{\nu+L}(\omega)&=n^\nu(\omega-LE)~\Leftrightarrow~\mu_{\nu+L}=\mu_\nu+LE,\\
    \end{split}
\end{equation}
which directly yields a similar relation for $\Gamma^{\nu,\mathrm{K}}_{ij}$. Finally, we assume that all terms in the Hamiltonian are strictly local, i.e., there exists an \(R\in\mathbb{N}\) such that
\begin{equation}
    \label{eq:support}
    \begin{split}
        h_{ij}&=\Gamma^{\nu,\mathrm{ret}}_{ij}=0\ ~\forall~ |i-j|\geq R,\\
        v_{ijkl}&=0 \hspace*{1.52cm}\forall~ \text{dist}(i,j,k,l)\geq R,
    \end{split}
\end{equation}
where $\tn{dist}(i,j,k,l)$ refers to the maximum of the pairwise distances $|i-j|$, $|i-k|$ etc. We stress that we do not impose any constraints on the size $L$ of the unit cell.

\begin{figure}
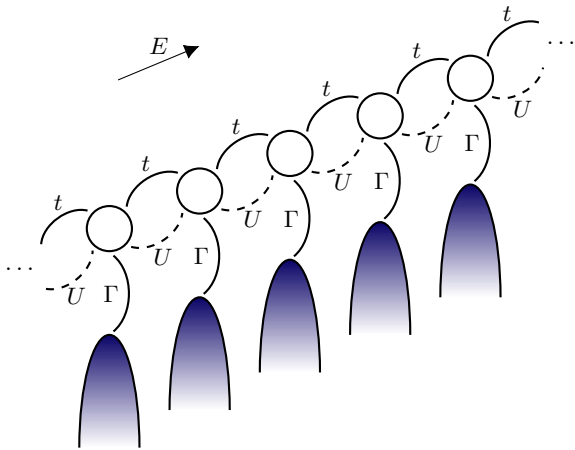

    \syst{-0.5}{1}
    \caption{ Pictorial representation of the model defined in Eq.~(\ref{eq:tbchain}) and used in Sec.~\ref{sec:application}. A tight-binding chain with a nearest-neighbor hopping \(t\) and a nearest-neighbor interaction \(U\) is subject to an in-plane electrical field $E$ and is coupled to wide-band reservoirs with a hybridization \(\Gamma\). In addition, we consider a finite staggered on-site potential of strength \(s\) (not depicted). }\label{fig:pic_chain}
\end{figure}

In Sec.~\ref{sec:application}, we will discuss the specific example of an interacting tight-binding chain coupled to zero-temperature wide-band reservoirs ($L=1, R=2$; see Fig.~\ref{fig:pic_chain}) governed by
\begin{equation}\label{eq:tbchain}
    \begin{split}
    &h_{00}=-U/2,~~ h_{01}=h_{10}=t,~~v_{0101}=U,\\
    &\Gamma^{0,\mathrm{ret}}_{0i}(\omega)= -\I \Gamma\delta_{i,0},~~ n^0(\omega)=\theta(-\omega).
    \end{split}
\end{equation}
All other components are uniquely defined by the symmetries of the system.
This model will serve as the main testbed for our method. In the limit $\Gamma=E=0$, the phase diagram can be computed analytically using the Bethe ansatz:\cite{giamarchi} The system is a gapless Luttinger liquid for \(U\leq 2t\) and a Mott insulator with a spontaneously-broken translational symmetry for \(U> 2t\), respectively. In the latter case, the ground-state is two-fold degenerate and features a charge-density wave (CDW). In order to break a potential ground-state degeneracy within a numerical method, we introduce a staggered potential $s$ that increases (decreases) the on-site energies of odd (even) sites within the chain by $s$ and therefore breaks translational symmetry. This increases the unit cell to \(L=2\).

\section{Green's functions}\label{sec:greens_functions}
We will now introduce Keldysh Green's functions, which are a key ingredient to diagrammatic methods such as the FRG formalism.\cite{Keldysh1965} The single-particle Green's functions in the stationary state can be written as
\begin{equation}
    G(\omega)=\begin{pmatrix}
        G^{11}(\omega) & G^{12}(\omega)\\
        G^{21}(\omega) & G^{22}(\omega)
    \end{pmatrix}=\begin{pmatrix}
        G^\mathrm{ret}(\omega) & G^\mathrm{K}(\omega)\\
        0 & G^\mathrm{adv}(\omega)
    \end{pmatrix}.
\end{equation}
The retarded component is given by 
\begin{equation}\label{eq:gr_from_self}
    \begin{split}
    G^\mathrm{ret}_{ij}(t,t')&=G^\mathrm{ret}_{ij}(t-t')=-\I\theta(t-t')\left\langle \left[c_j^\dagger(t'),c_i(t)\right]_+\right\rangle,\\
    G^\mathrm{ret}_{ij}(\omega)&=\int_{-\infty}^\infty \mathrm{d}t \mathrm{e}^{\I\omega t} G^\mathrm{ret}_{ij}(t)
    =G^\mathrm{adv}_{ji}(\omega)^*,
    \end{split}
\end{equation}
and can be related to the non-interacting retarded Green's function $g^\mathrm{ret}(\omega)$ by virtue of the Dyson equation:
\begin{equation}\label{eq:dysonret}\begin{split}
        G^\mathrm{ret}(\omega) & =\frac{1}{g^\mathrm{ret}(\omega)^{-1}-\Sigma^\mathrm{ret}(\omega)}, \\
        g^\mathrm{ret}(\omega)&=\frac{1}{\omega-h- \sum_{\nu\in \mathbb{Z}} \Gamma^{\nu,\mathrm{ret}}(\omega)},
\end{split}\end{equation}
where the self-energy $\Sigma^\mathrm{ret}$ is associated with the two-particle interaction $v_{ijkl}$. The Keldysh Green's function is defined as 
\begin{equation}
    \begin{split}
        G^\mathrm{K}_{ij}(t-t')&=\I\left[ \left\langle c_{j}^\dagger(t') c_i(t)\right\rangle-\left\langle c_i(t) c_{j}^\dagger(t') \right\rangle\right],\\
    G^\mathrm{K}(\omega)&=\int_{-\infty}^\infty \mathrm{d} t \mathrm{e}^{i\omega t} G^\mathrm{K}(t),
\end{split}
\end{equation}
and the corresponding Dyson equation reads
\begin{equation}\label{eq:gk_from_self}
    \begin{split}
         G^\mathrm{K}& = G^\mathrm{ret}[ (g^\mathrm{ret})^{-1} g^\mathrm{K}(g^\mathrm{adv})^{-1}+  \Sigma^\mathrm{K}] G^\mathrm{adv}\\
         & = G^\mathrm{ret}\Big[\sum_{\nu\in\mathbb{Z}}\Gamma^{\nu,\mathrm{K}} +  \Sigma^\mathrm{K}\Big] G^\mathrm{adv},
    \end{split}
\end{equation}
where we have used that
\begin{equation}\label{eq:gk}
g^\mathrm{K}=g^\mathrm{ret}\sum_{\nu\in\mathbb{Z}}\Gamma^{\nu,\mathrm{K}} g^\mathrm{adv}. 
\end{equation}
All quantities in Eqs.~(\ref{eq:gr_from_self}) and (\ref{eq:gk_from_self}) are matrices defined by two single-particle indices. To simplify the notation, we will frequently employ multi-indices \(1=(i_1, \alpha_1)\) that include both this single-particle index $i_1$ as well as the Keldysh index $\alpha_1\in\{1,2\}$. The frequency-dependence will still be written out explicitly.

If the entire system is in an equilibrium configuration described by the (Fermi) distribution function \(n(\omega)\), the Green's functions obey the fluctuation-dissipation theorem:
\begin{equation}\label{eq:fluc_dis}
    G^\mathrm{K}(\omega)=\left[1-2n(\omega)\right] \left[ G^\mathrm{ret}(\omega)-G^\mathrm{adv}(\omega)\right].
\end{equation}
The FRG approximation we introduce in Sec.~\ref{sec:fRG} preserves this symmetry in the equilibrium limit, which is essential in order to avoid unphysical, anomalous heating effects.

The symmetry described by Eqs.~(\ref{eq:shift_sym}) and (\ref{eq:shift_sym_res}) translates directly to non-interacting Green's function $g$,
\begin{equation}
        g_{(1'+L)(1+L)}(\omega)=g_{1'1}(\omega-LE),
\end{equation}
where $1+L$ denotes a shift of the single-particle index, $1+L=(i_1+L,\alpha_1)$. This is a direct consequence of the Dyson equations (\ref{eq:dysonret}) and (\ref{eq:gk}). It follows from diagrammatic arguments (an expansion into an infinite perturbation series) that the \textit{exact} self-energy $\Sigma$ and thus also the full Green function $G^\Lambda$ (see, e.g., the Dyson equation) inherit this symmetry:
\begin{equation}\label{eq:shift_sym_self_gf}
    \begin{split}
        \Sigma_{(1'+L)(1+L)}(\omega)&=\Sigma_{1'1}(\omega-LE),\\
        G_{(1'+L)(1+L)}(\omega)&=G_{1'1}(\omega-LE).\\
    \end{split}
\end{equation}

\section{Computing Green's functions in an infinite system}\label{sec:iterative_gf}

In this section, we discuss how to compute the retarded and Keldysh Green's function of an infinite system under the assumption that the corresponding self-energies are known (e.g., from an FRG calculation). This cannot be done straightforwardly but requires an iterative algorithm, which we will now present. Our algorithm does not involve any additional approximations but is (numerically) exact. More importantly, it is not specifically tailored to the methods of this paper but is applicable in a completely general setting.

In the following, we will assume that i) our system fulfills the translation symmetry of Eqs.~(\ref{eq:shift_sym}) and (\ref{eq:shift_sym_res}), that ii) the single-particle Hamiltonian $h$, the self-energies $\Sigma^{\tn{ret,K}}$, and the reservoir coupling $\Gamma_{ij}^{\nu, \textnormal{ret}}$ are of limited range $N$ where \(L\) evenly divides \(N\), i.e.,
\begin{equation}\label{eq:support2}
h_{ij}=\Gamma_{ij}^{\nu, \textnormal{ret}}=\Sigma^{\tn{ret,K}}_{ij}=0 ~~~\tn{if}~ |i-j|\geq N, 
\end{equation}
and that iii) the Green's functions $G^\textnormal{ret,K}_{ij}$ are only needed for $|i-j|<N$ (this will be the case in the FRG approach introduced in the next section).

In our concrete example of the tight-binding chain [see Eq.~(\ref{eq:tbchain})], we have $h_{ij}=\Gamma_{ij}^{\nu, \textnormal{ret}}=0$ for $|i-j|\geq 2$. In Sec.~\ref{sec:fRG}, we will show that our FRG approximation to the self-energy fulfills Eq.~(\ref{eq:support2}) and that only Green's function with $|i-j|< 3M$ enter into the flow equations due to the approximations made in Eqs.~(\ref{eq:chan_decomp}) and~(\ref{eq:truncM}). Thus, we would have $N=3M$ in this case.

A similar recursive algorithm has been put forward in Ref.~\onlinecite{Neumayer2015} in the context of a cluster perturbation theory calculation. We generalize those ideas by, e.g., allowing for an arbitrary, finite-range self-energy.

\subsection{Notation}\label{ssec:notation}
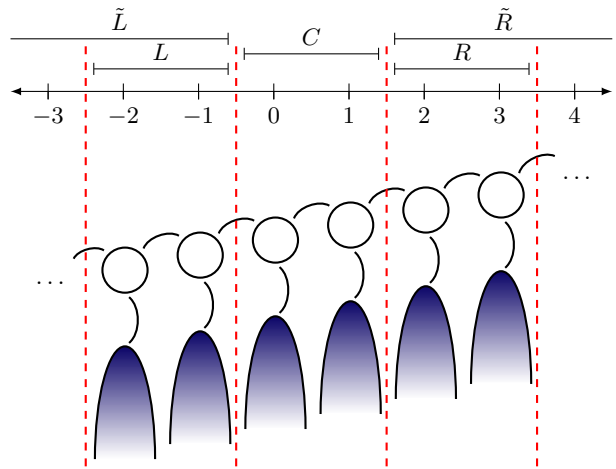
\begin{figure}
    \begin{tikzpicture}
\draw[latex-latex] (-3.5,0) -- (4.5,0) ;
\foreach \x in  {-3,-2,...,4}
\draw[shift={(\x,0)},color=black] (0pt,3pt) -- (0pt,-3pt);
\foreach \x in {-3,-2,...,4}
\draw[shift={(\x,0)},color=black] (0pt,0pt) -- (0pt,-3pt) node[below] 
{$\x$};
\draw[|-|] (-0.4,0.5) -- node[above]{$C$}         ( 1.4,0.5) ;
\draw[|-|] (-2.4,0.3) -- node[above]{$L$}         (-0.6,0.3) ;
\draw[|-|] ( 1.6,0.3) -- node[above]{$R$}         ( 3.4,0.3) ;
\draw[ -|] (-3.5,0.7) -- node[above]{$\tilde{L}$} (-0.6,0.7) ;
\draw[|- ] ( 1.6,0.7) -- node[above]{$\tilde{R}$} ( 4.5,0.7) ;
    \def\efi{0.2}
    \coordinate(A0) at (-4.2,-3.0);
    \foreach \a in {1,2,...,8}{
        \if\a1%
            \node[circle, minimum size=0.6cm, above right=\efi*\a and 1.0*\a of A0, thick](A\a){$\dots$};
        \else
            \if\a8%
                \node[circle, minimum size=0.6cm, above right=\efi*\a and 1.0*\a of A0, thick](A\a){$\dots$};
            \else
                \node[draw, circle, minimum size=0.6cm, above right= \efi*\a and 1.0*\a of A0, thick](A\a){};
            \fi
        \fi;
    }

    \foreach[evaluate=\a as \an using int(\a+1)] \a in {1,2,...,7}
        \path[]
            (A\a.north east) edge [
              text=black,
              thick,
              shorten <=2pt,
              shorten >=2pt,
              bend left=50
              ] node[above]{
                           } (A\an.north west);

       \draw[red, dashed, thick] (-2.5,0.6) to (-2.5,-5);
       \draw[red, dashed, thick] (-0.5,0.6) to (-0.5,-5);
       \draw[red, dashed, thick] ( 1.5,0.6) to ( 1.5,-5);
       \draw[red, dashed, thick] ( 3.5,0.6) to ( 3.5,-5);

   \foreach \a in {2,3,...,7}{
          \coordinate[below=2.2cm of A\a](r\a);
          \draw[thick, shading=axis, bottom color=white, top color=custBlue] (r\a)++(-0:0.4 and 1.5) arc(-0:180:0.4 and 1.5);
        \path[]
            (r\a)++(90:0.4 and 1.5) edge [
              text=black,
              thick,
              shorten <=2pt,
              shorten >=2pt,
              bend right=50
              ] node[left]{
                           } (A\a.south);
    }
\end{tikzpicture}
\caption{Illustration of the decomposition of the system used in Sec.~\ref{ssec:notation} for the example of a simple tight-binding chain with \(N=2\).}
\label{fig:splitsys}
\end{figure}

For the rest of this section, we introduce the following notation for the single-particle indices (see Fig.~\ref{fig:splitsys}):
\begin{equation}\begin{split}
i\in\lli ~~& \Leftrightarrow~~ \hspace*{1.22cm} i \leq -1, \\
i\in\li ~~& \Leftrightarrow~~ -N \leq  i \leq -1, \\
i\in\ci ~~& \Leftrightarrow~~ \hspace*{0.58cm}0 \leq  i \leq N-1, \\
i\in\ri ~~& \Leftrightarrow~~ \hspace*{0.43cm}N \leq  i \leq 2N-1, \\
i\in\rri ~~& \Leftrightarrow~~\hspace*{0.43cm}N \leq  i , \\
\end{split}\end{equation}
with the implicit understanding that, e.g.,  $G^\textnormal{ret}_{\ci\li}(\omega)$ refers to the retarded Green's function $G^\textnormal{ret}_{(i\in\ci)(j\in\li)}(\omega)$, which is a matrix of size $N\times N$. The same convention is used for the self-energy as well as for all other quantities carrying two single-particle indices. We employ an Einstein convention for summations, e.g.,
\begin{equation}\begin{split}
G^\textnormal{ret}_{\ci\li}(\omega) \Sigma^\textnormal{K}_{\li\li}(\omega) =
\sum_{j\in\li} \left[G^\textnormal{ret}_{ij} (\omega)\Sigma^\textnormal{K}_{jk}(\omega)\right]_{i\in\ci, k \in\li}.
\end{split}\end{equation}
We finally note that using our notation, the translation symmetry in Eq.~(\ref{eq:shift_sym_self_gf}) takes the form
\begin{equation}\label{eq:iter_translation}
 G^\textnormal{ret,K}_{\li\li}(\omega-NE) = G^\textnormal{ret,K}_{\ci\ci}(\omega) = G^\textnormal{ret,K}_{\ri\ri}(\omega+NE),
\end{equation}
and likewise for the self-energy.

\subsection{Retarded Green's function}

We first discuss how one can obtain the retarded part of the Green's function. We need to invert a matrix which by construction has the following form:
\begin{equation}\label{eq:iter_heff}
    \omega - h-\sum_{\nu\in\mathbb{Z}}\Gamma^{\nu,\mathrm{ret}}-\Sigma^\mathrm{ret}= T+D,
\end{equation}
where a block structure is defined by $T$ and $D$ as follows:
\begin{equation}
    T+D=
  \left(\begin{array}{cc|c|cc}
        &         &     0 &      0 &0 \\
      \multicolumn{2}{c|}{\smash{\raisebox{.5\normalbaselineskip}{$D_{\lli\lli}$}}}
        & T_{\li\ci} &      0 & 0 \\
      \hline \\[-\normalbaselineskip]
      0 &  T_{\ci\li} &     D_{\ci\ci} & T_{\ci\ri} & 0\\
      \hline\\[-\normalbaselineskip]
       0&     0   &     T_{\ri\ci} &       & \\
      
      0 &      0 & 0 & \multicolumn{2}{|c}{\smash{\raisebox{.5\normalbaselineskip}{$D_{\rri\rri}$}}} \\
    \end{array}\right).
\end{equation}
$D_{\ci\ci}$, $T_{\ci\li}$, $T_{\li\ci}$, $T_{\ci\ri}$, and $T_{\ri\ci}$ are matrices of size $N\times N$. Note that in general \(T_{\li\ci}\neq T_{\ci\li}^\dag\) due to the inclusion of the self-energy. From now on, we will often omit the frequency dependence to improve readability.

A crucial ingredient is that the inverse of a block matrix is given by
\begin{equation}\label{eq:iter_matinv}\begin{split}
\begin{pmatrix} V & W \\ X & Y \end{pmatrix}^{-1}
&= \begin{pmatrix} V^{-1} + V^{-1}WY_IXV^{-1} & -V^{-1}WY_I \\ -Y_IXV^{-1} & Y_I \end{pmatrix} \\[1ex]
Y_I&= (Y-XV^{-1}W)^{-1},
\end{split}\end{equation}
or equivalently
\begin{equation}\label{eq:iter_matinv2}\begin{split}
\begin{pmatrix} V & W \\ X & Y \end{pmatrix}^{-1}
&= \begin{pmatrix} V_I & -V_IWY^{-1} \\ -Y^{-1}XV_I & Y^{-1}+Y^{-1}XV_IWY^{-1} \end{pmatrix} \\[1ex]
V_I&= (V-WY^{-1}X)^{-1}.
\end{split}\end{equation}
By successively applying Eq.~\eqref{eq:iter_matinv} and (\ref{eq:iter_matinv2}), one can prove that
    \begin{equation}\label{eq:iter_matinv3}
        \begin{split}
            \left[\left(\begin{array}{c|cc} N & U & 0 \\ \hline V & W & X \\ 0 & Y & Z \end{array}\right)^{-1}\right]_{22}&=
        \left[\begin{pmatrix} W-VN^{-1}U & X\\ Y & Z\end{pmatrix}^{-1}\right]_{11}\\
        &= \left(W-VN^{-1}U-XZ^{-1}Y\right)^{-1},
    \end{split}
    \end{equation}
where in the first step we identified four blocks as indicated on the lhs.

Per our assumption, the retarded Green's function $G^\tn{ret}_{ij}$ is only needed for $|i-j|<N$; it is thus sufficient to determine $G^{\tn{ret}}_{\ci\ci}$. If we apply Eq.~(\ref{eq:iter_matinv3}) to Eq.~(\ref{eq:iter_heff}) and use that \(T_{\ci(\lli\setminus\li)}=0\), we obtain
\begin{equation}\label{eq:iter_greta}\begin{split}
& G^{\tn{ret}}_{\ci\ci}(\omega) = \frac{1}{D_{\ci\ci}-T_{\ci\li}[D^{-1}]_{\li\li}T_{\li\ci} -T_{\ci\ri}[D^{-1}]_{\ri\ri}T_{\ri\ci} }.
\end{split}\end{equation}
In order to solve Eq.~(\ref{eq:iter_greta}), we need to determine the objects $[D^{-1}]_{\li\li}$ as well as $[D^{-1}]_{\ri\ri}$, i.e., we need to calculate the first and last block of the inverse of the two matrices $D_{\lli\lli}$ and $D_{\rri\rri}$ associated with semi-infinite systems. This can be achieved iteratively by exploiting translation-invariance, which we will now discuss.

\subsubsection*{Iterative algorithm for the auxiliary Green's function}

For notational simplicity, we define an auxiliary Green's function
\begin{equation}
\calrLR:=D^{-1} 
\end{equation}
as the inverse of the matrix in Eq.~(\ref{eq:iter_heff}) for $T_{\li\ci}=T_{\ci\li}=T_{\ri\ci}=T_{\ci\ri}=0$ (which becomes block diagonal in this case). For later use, we introduce similar objects $\calrL$ and $\calrR$ as the inverse of Eq.~(\ref{eq:iter_heff}) where only $T_{\li\ci}=T_{\ci\li}=0$ and $T_{\ri\ci}=T_{\ci\ri}=0$, respectively. The advanced components are defined as $\calaLR:=[\calrLR]^\dagger$, $\calaL:=[\calrL]^\dagger$, and $\calaR:=[\calrR]^\dagger$. We note that while in the presence of finite interactions $\Sigma^\tn{ret}\neq0$, these are no longer physical Green's functions of the underlying Hamiltonian, they inherit all of its symmetries such as translation-invariance:
\begin{equation}\label{eq:iter_translation2}\begin{split}
\calrL_{\li\li}(\omega-NE) &= \calrLR_{\li\li}(\omega-NE) = \calrR_{\ci\ci}(\omega),\\
\calrR_{\ri\ri}(\omega+NE) &= \calrLR_{\ri\ri}(\omega+NE) = \calrL_{\ci\ci}(\omega).
\end{split}\end{equation}
The first line is a direct consequence of Eqs.~(\ref{eq:shift_sym}), (\ref{eq:shift_sym_res}), and (\ref{eq:shift_sym_self_gf}) combined with the fact that the last $N\times N$ block ($\li$) of an isolated $\lli$ system is, up to a shift in energy, identical to the last $N\times N$ block ($\ci$) of a system where $\rri$ is removed (the second line follows similarly). If we use Eq.~(\ref{eq:iter_translation2}), we can now set up a recursion relation to determine $\calrLR_{\li\li}$:
\begin{equation}\label{eq:iter_gf2}\begin{split}
& \calrLR_{\li\li}(\omega-NE) = \calrR_{\ci\ci}(\omega) \\
    = &~ \frac{1}{D_{\ci\ci}(\omega)-T_{\ci\li}(\omega)\calrLR_{\li\li}(\omega)T_{\li\ci}(\omega)},
\end{split}\end{equation}
where we have applied Eq.~(\ref{eq:iter_matinv}) to Eq.~(\ref{eq:iter_heff}) with $T_{\ci\ri}=T_{\ri\ci}=0$. A similar expression can be derived for $\calrLR_{\ri\ri}(\omega)$.

For \(E=0\), Eq.~(\ref{eq:iter_gf2}) is local in \(\omega\) and easily solved using a self-consistency loop. 
At finite electric field, however, this equation couples Green's functions at different frequencies. One can solve it by using \(\lim_{\omega\rightarrow \pm \infty} \calrR(\omega)=0\) as an initial condition; in practice, it is sufficient to set \(\calrR(\pm\Omega)=0\), where \(\Omega\) far exceeds all other energy scales. Eq.~\eqref{eq:iter_gf2} can then be used to successively calculate the auxiliary Green's function on a discrete grid of frequencies. How to do this in practice is outlined in Appendix~\ref{sec:sol_self_cons}.

\subsection{Keldysh Green's function}

Next, we illustrate how to compute the Keldysh Green's function $G^\tn{K}_{\ci\ci}(\omega)$. The Dyson equation (\ref{eq:gk_from_self}) takes the form
\begin{equation}
\begin{split}
& G^\tn{K}_{\ci\ci}(\omega) = \sum_{\si,\si'=\lli,\ci,\rri}G^\tn{ret}_{\ci\si}(\omega)
\Big[\sum_{\nu\in\mathbb{Z}}\Gamma^{\nu,\mathrm{K}}_{\si\si'} +  \Sigma^\mathrm{K}_{\si\si'}\Big] G^\tn{adv}_{\si'\ci}(\omega).
\end{split}
\end{equation}
If we employ the lower-left component of Eq.~(\ref{eq:iter_matinv}),
\begin{equation}\label{eq:iter_dyson2a}\begin{split}
G^{\tn{ret}}_{\ci\lli} &= - G^{\tn{ret}}_{\ci\ci}T_{\ci\li}[D^{-1}]_{\li\lli}= - G^{\tn{ret}}_{\ci\ci}T_{\ci\li}\calrLR_{\li\lli}, \\
G^{\tn{ret}}_{\ci\rri} &= - G^{\tn{ret}}_{\ci\ci}T_{\ci\ri}[D^{-1}]_{\ri\rri}= - G^{\tn{ret}}_{\ci\ci}T_{\ci\ri}\calrLR_{\ri\rri},
\end{split}\end{equation}
the Dyson equation can be simplified as follows:
\begin{equation}\label{eq:iter_gk1}
\begin{split}
 G^\tn{K}_{\ci\ci} = G^\tn{ret}_{\ci\ci}\Big[ \tsk_{\ci\ci} - & T_{\ci\li}\calrLR_{\li\li}\tsk_{\li\ci} - \tsk_{\ci\li} \calaLR_{\li\li} T^\dagger_{\li\ci} \\
 -&T_{\ci\ri}\calrLR_{\ri\ri}\tsk_{\ri\ci}  - \tsk_{\ci\ri} \calaLR_{\ri\ri} T^\dagger_{\ri\ci} \\
  +& T_{\ci\li}\calkLR_{\li\li} T^\dagger_{\li\ci}+ T_{\ci\ri}\calkLR_{\ri\ri}T^\dagger_{\ri\ci}\Big]G^\tn{adv}_{\ci\ci},
\end{split}
\end{equation}
where we have defined
\begin{equation}\label{eq:tsk}
\tsk:=\sum_{\nu\in\mathbb{Z}}\Gamma^{\nu,\mathrm{K}} +  \Sigma^\mathrm{K}
\end{equation}
as well as the auxiliary Green's functions
\begin{equation}\begin{split}
\calkLR_{\li\li}(\omega) &\coloneqq \calrLR_{\li\lli}(\omega)\tsk_{\lli\lli}(\omega)\calaLR_{\lli\li}(\omega), \\
\calkLR_{\ri\ri}(\omega) &\coloneqq \calrLR_{\ri\rri}(\omega)\tsk_{\rri\rri}(\omega)\calaLR_{\rri\ri}(\omega).
\end{split}\end{equation}
The latter are the only unknown quantities in Eq.~(\ref{eq:iter_gk1}); $G^\tn{ret}_{\ci\ci}$, $\calrLR_{\li\li}$, and $\calrLR_{\ri\ri}$ have already been calculated in the previous section. In Eq.~(\ref{eq:iter_gk1}), we have employed that $\tsk_{\lli\rri}=\tsk_{\ci(\lli\setminus\li)}=\tsk_{\ci(\rri\setminus\ri)}=0$ holds per our assumption. We will now discuss how $\calkLR_{\li\li}$ and $\calkLR_{\li\li}$ can be computed iteratively by exploiting translation-invariance.

\subsubsection*{Iterative algorithm for the auxiliary Green's function}

In order to compute $\calkLR_{\li\li}$ and $\calkLR_{\ri\ri}$, we will use the relation
\begin{equation}\label{eq:iter_dyson2b}\begin{split}
\calrR_{\ci\lli} &= - \calrR_{\ci\ci}T_{\ci\li}\calrLR_{\li\lli}, \\
\calrL_{\ci\rri} &= - \calrL_{\ci\ci}T_{\ci\ri}\calrLR_{\ri\rri},
\end{split}\end{equation}
which is analogous to Eq.~(\ref{eq:iter_dyson2a}) and follows by applying Eq.~(\ref{eq:iter_matinv}) to Eq.~(\ref{eq:iter_heff}) with $T_{\ci\ri}=T_{\ri\ci}=0$. Moreover, we exploit translation-invariance for $\tsk$ [see Eq.~(\ref{eq:shift_sym_self_gf})] as well as for ${\cal G}^{\tn{ret,adv}}$:
\begin{equation}\label{eq:iter_translation3}\begin{split}
\calrLR_{\li\lli}(\omega-NE) &= \calrR_{\ci(\ci\cup\lli)}(\omega),\\
\calrLR_{\ri\rri}(\omega+NE) &= \calrL_{\ci(\ci\cup\rri)}(\omega),
\end{split}\end{equation}
which can be derived in analogy to Eq.~(\ref{eq:iter_translation2}). This yields
\begin{equation}\label{eq:iter_gk2}
\begin{split}
&~ \calkLR_{\li\li}(\omega-NE)\\
=\hspace*{0.1cm} &~  \calrLR_{\li\lli}(\omega-NE)\tsk_{\lli\lli}(\omega-NE)\calaLR_{\lli\li}(\omega-NE)\\
\stackrel{\tn{(\ref{eq:iter_translation3})}}{=}&~ \calrR_{\ci(\ci\cup\lli)}(\omega)\tsk_{(\ci\cup\lli)(\ci\cup\lli)}(\omega)\calaR_{(\ci\cup\lli)\ci}(\omega)\\
\stackrel{\tn{(\ref{eq:iter_dyson2b})}}{=} &~ \calrR_{\ci\ci}\Big[\tsk_{\ci\ci} - T_{\ci\li}\calrLR_{\li\li}\tsk_{\li\ci}  - \tsk_{\ci\li}\calaLR_{\li\li} T^\dagger_{\li\ci} \\
&\hspace*{1.2cm} + T_{\ci\li}\underbrace{\calrLR_{\li\lli}\tsk_{\lli\lli}\calaLR_{\lli\li}}_{=\calkLR_{\li\li}(\omega)} T^\dagger_{\li\ci}\Big]\calaR_{\ci\ci}.
\end{split}
\end{equation}
In the last line, all quantities carry a frequency argument $\omega$, which we have omitted to improve readability. $\calkLR_{\ri\ri}$ follows similarly. This equation has the same form as Eq.~(\ref{eq:iter_gf2}) and can be solved either self-consistently (if \(E=0\)) or successively by utilizing \(\calkLR(\omega)\to 0\) for \(\omega\to \pm\infty\) (if \(E\neq 0\), see Appendix~\ref{sec:sol_self_cons}).

\section{Functional renormalization group approach} \label{sec:fRG}
The functional renormalization group is an implementation of the RG idea on the level of correlation functions.\cite{Metzner2012} It sets up flow equations for the self-energy as well as for higher-order vertex functions with respect to a flow parameter $\Lambda$ introduced as an infrared cutoff within the non-interacting Green's functions $g^{\mathrm{ret},\Lambda}(\omega)$ and $g^{\mathrm{K},\Lambda}(\omega)$. A detailed description to this method can be found in Refs.~\onlinecite{Metzner2012,kopietzBook}.

From now on, we focus solely on the case that $H_\mathrm{res}^\nu$ describes zero-temperature wide-band reservoirs with a  frequency independent hybridization that are coupled uniformly to the chain:
\begin{equation}\label{eq:frgcutoff}
\Gamma_{ij}^{\nu,\mathrm{ret}}(\omega)=-\I\Gamma\delta_{i,j}\delta_{i,\nu}.
\end{equation}
In this case, it is convenient to use the single scale \(\Gamma\) as the flow parameter \(\Lambda=\Gamma\).\cite{Jakobs2010b} The advantage of this approach is its physical interpretation -- intermediate results during the solution of the flow equations can simply be viewed as physical results at a stronger coupling.

We will now set up a full-fledged second-order Keldysh FRG implementation for a one-dimensional chain that is translation-invariant up to shifts in energy. A variety of different first-order Keldysh FRG calculations can be found in the literature,\cite{Gezzi2007,Jakobs2007,Karrasch2010,Kennes2012} but second-order FRG schemes have so far been developed exclusively for single-impurity models\cite{Karrasch2008,Jakobs2010b} or for chains which are in thermal equilibrium.\cite{Heyder2014,Sbierski2017,Markhof2018,Weidinger2019}

The only key approximation in our scheme is the so-called channel decomposition of the vertex flow equation (see Sec.~\ref{sec:frg_channel}), which has been widely applied in equilibrium FRG calculations.\cite{Karrasch2008,Sbierski2017,Markhof2018,Weidinger2019} Moreover, we will assume that all vertex functions have a limited support of range $M$ (see Sec.~\ref{sec:frg_support}), which will serve as our key numerical control parameter. The original second-order flow equations (which only assume the channel decomposition) are recovered in the limit $M\to\infty$, and we will demonstrate that convergence in $M$ can be reached in all practical applications.

\subsection{Flow equations}
The flow equation for the self-energy reads
\begin{equation}
    \label{eq:ford}
    \partial_\Lambda\Sigma^\Lambda_{1'1}(\omega)=-\frac{\I}{2\pi} \int d\Omega\sum_{22'} \gamma^\Lambda_{1'2'12}(\omega, \Omega,\omega, \Omega) S^\Lambda_{22'}(\Omega).
\end{equation}
The single-scale propagator is given by
\begin{equation}\begin{split}\label{eq:single_scale}
S^\Lambda(\omega) & =-G^\Lambda(\omega) \{\partial_\Lambda [g^{\Lambda}(\omega)^{-1}]\}G^\Lambda(\omega) \\& = \partial_\Lambda^* G^\Lambda(\omega),
\end{split}\end{equation}
where $\partial_\Lambda^*$ indicates a derivative that acts only on the explicit $\Lambda$-dependence of the cutoff (but not on $\Sigma^\Lambda$). The quantity \(\gamma\) denotes the one-particle irreducible two-particle vertex function; it preserves energy conservation due to the time-translation invariance of the system, and its frequency-dependence can thus be parametrized via
\begin{equation}
\gamma^\Lambda_{1'2'12}(\omega_{1'},\omega_{2'},\omega_1, \omega_2)=\gamma^\Lambda_{1'2'12}(\Pi,X,\Delta)
\end{equation}
with the coordinates
\begin{equation}
    \label{eq:freq_trafo}
    \begin{split}
        \Pi&=\omega_1+\omega_2=\omega_{1'}+\omega_{2'},\\
        X& =\omega_{2'}-\omega_1=\omega_{2}-\omega_{1'},\\
        \Delta& =\omega_{1'}-\omega_1=\omega_{2}-\omega_{2'}.
    \end{split}
\end{equation}
Using this notation, the flow equation for $\gamma$ takes the form
\begin{widetext}
\begin{equation}
    \label{eq:sord}
    \begin{split}
        \partial_\Lambda \gamma^\Lambda_{1'2'12}(\Pi,X,\Delta)=&\frac{\I}{2\pi}\int d\Omega\sum_{33'44'}\\[2ex]
        &\hspace{-2cm}\gamma^\Lambda_{1'2'34}\left(\Pi, \Omega+\frac{X-\Delta}{2}, \Omega-\frac{X-\Delta}{2}\right)S^\Lambda_{33'}\left(\frac{\Pi}{2}-\Omega\right)G^\Lambda_{44'}\left(\frac{\Pi}{2}+\Omega\right)\gamma^\Lambda_{3'4'12}\left(\Pi,\frac{X+\Delta}{2}+\Omega, \frac{X+\Delta}{2}-\Omega\right)\\[2ex]
       &\hspace{-2.4cm}+\gamma^\Lambda_{1'4'32}\left(\frac{\Pi+\Delta}{2}+\Omega, X, \frac{\Pi+\Delta}{2}-\Omega\right)\biggl[S^\Lambda_{33'}\left(\Omega-\frac{X}{2}\right)G^\Lambda_{44'}\left(\Omega+\frac{X}{2}\right)+\\[2ex]
        &\hspace{3.86cm}G^\Lambda_{33'}\left(\Omega-\frac{X}{2}\right)S^\Lambda_{44'}\left(\Omega+\frac{X}{2}\right)\biggr]\gamma^\Lambda_{3'2'14}\left(\Omega+\frac{\Pi-\Delta}{2}, X, \Omega- \frac{\Pi-\Delta}{2}\right)\\[2ex]
        &\hspace{-2.4cm}-\gamma^\Lambda_{1'3'14}\left(\Omega+\frac{\Pi-X}{2}, \Omega-\frac{\Pi-X}{2}, \Delta\right)\biggl[S^\Lambda_{33'}\left(\Omega-\frac{\Delta}{2}\right)G^\Lambda_{44'}\left(\Omega+\frac{\Delta}{2}\right)+\\[2ex]
        &\hspace{3.91cm}G^\Lambda_{33'}\left(\Omega-\frac{\Delta}{2}\right)S^\Lambda_{44'}\left(\Omega+\frac{\Delta}{2}\right)\biggr]\gamma^\Lambda_{4'2'32}\left(\frac{\Pi+X}{2}+\Omega, \frac{\Pi+X}{2}-\Omega,\Delta\right)\\[2ex]
        &+\mathcal{O}(U^3),
    \end{split}
\end{equation}
\end{widetext}
where we already truncated the otherwise infinite hierarchy of differential equations by neglecting the flow of the three-particle vertex. This approximation is controlled in a perturbative sense and all terms neglected are at least of \(\mathcal{O}\left(U^3\right)\).

The flow equations (\ref{eq:ford}) and (\ref{eq:sord}) need to be complemented by an initial condition. When the coupling to the reservoirs is large ($\Lambda\to\infty$), the vertex functions can be obtained analytically:
\begin{equation}
    \label{eq:ini_cond}
    \begin{split}
        \Sigma^{\mathrm{ret}, \Lambda\to\infty}_{i'i}&=\frac{1}{2}\sum_j v_{i'jij},~
        \Sigma^{\mathrm{K},\Lambda\to\infty}_{i'i}=0,~
        \gamma^{\Lambda\to\infty}_{1'2'12}=\bar v_{1'2'12},
    \end{split}
\end{equation}
where we introduced the Keldysh-space version of the two-particle interaction
\begin{equation}
    \bar v_{1'2'12}=\begin{cases} \frac{1}{2} v_{i_{1'} i_{2'} i_1 i_2} & \alpha_{1'}+\alpha_{2'}+\alpha_1+\alpha_2\ \text{odd}\\
            0 & \text{otherwise.}\end{cases}
\end{equation}
The initial value of the retarded self-energy is frequency-independent and can therefore be absorbed into the non-interacting Hamiltonian $h$.

\subsection{Channel decomposition}
\label{sec:frg_channel}

The vertex flow equation (\ref{eq:sord}) depends on three independent frequencies and is thus difficult to tackle numerically. Hence, we need to resort to an additional approximation, the so-called channel decomposition.\cite{Karrasch2008} We make the following ansatz for $\gamma$:
\begin{equation}
    \label{eq:chan_decomp}
    \begin{split}
        \gamma^\Lambda_{1'2'12}(\Pi,X,\Delta)=&~\bar v_{1'2'12}+\gamma^{\mathrm{p},\Lambda}_{1'2'12}(\Pi)\\
                                   &+\gamma^{\mathrm{x},\Lambda}_{1'2'12}(X)+\gamma^{\mathrm{d},\Lambda}_{1'2'12}(\Delta),
    \end{split}
\end{equation}
and assume that (i) the flow equation for $\gamma^{\mathrm{p},\Lambda}$, $\gamma^{\mathrm{x},\Lambda}$, and $\gamma^{\mathrm{d},\Lambda}$ is given by the first, second, and third term of Eq.~(\ref{eq:sord}), respectively, and that (ii) each channel is only fed back into its own flow equation. This yields
\begin{widetext}
\begin{equation}
    \label{eq:chan_decomp_flow}
    \begin{split}
        \partial_\Lambda \gamma^{\mathrm{p},\Lambda}_{1'2'12}(\Pi)=&\frac{\I}{2\pi}\int d\Omega\sum_{33'44'}\bar\gamma^{\mathrm{p},\Lambda}_{1'2'34}\left(\Pi\right)S^\Lambda_{33'}\left(\frac{\Pi}{2}-\Omega\right)G^\Lambda_{44'}\left(\frac{\Pi}{2}+\Omega\right)\bar\gamma^{\mathrm{p},\Lambda}_{3'4'12}\left(\Pi\right),\\
        \partial_\Lambda \gamma^{\mathrm{x},\Lambda}_{1'2'12}(X)=&\frac{\I}{2\pi}\int d\Omega\sum_{33'44'}\bar\gamma^{\mathrm{x},\Lambda}_{1'4'32}\left( X\right)\biggl[S^\Lambda_{33'}\left(\Omega-\frac{X}{2}\right)G^\Lambda_{44'}\left(\Omega+\frac{X}{2}\right)+G^\Lambda_{33'}\left(\Omega-\frac{X}{2}\right)S^\Lambda_{44'}\left(\Omega+\frac{X}{2}\right)\biggr]\bar\gamma^{\mathrm{x},\Lambda}_{3'2'14}\left(X\right),\\
        \partial_\Lambda \gamma^{\mathrm{d},\Lambda}_{1'2'12}(\Delta)=&\frac{-\I}{2\pi}\int d\Omega\sum_{33'44'}\bar\gamma^{\mathrm{d},\Lambda}_{1'3'14}\left(\Delta\right)\biggl[S^\Lambda_{33'}\left(\Omega-\frac{\Delta}{2}\right)G^\Lambda_{44'}\left(\Omega+\frac{\Delta}{2}\right)+G^\Lambda_{33'}\left(\Omega-\frac{\Delta}{2}\right)S^\Lambda_{44'}\left(\Omega+\frac{\Delta}{2}\right)\biggr]\bar\gamma^{\mathrm{d},\Lambda}_{4'2'32}\left(\Delta\right),
    \end{split}
\end{equation}
\end{widetext}
with \(\bar \gamma^{\alpha,\Lambda}=\bar v+\gamma^{\alpha,\Lambda}\), \(\alpha=\mathrm{p,x,d}\). The initial condition reads $\gamma^{\alpha,\Lambda\to\infty}_{1'2'12}=0$. The self-energy flow equation (\ref{eq:ford}) now takes the form
\begin{equation}\label{eq:ford2}
\begin{split}
    \partial_\Lambda&\Sigma^\Lambda_{1'1}(\omega) =-\frac{\I}{2\pi} \int d\Omega\sum_{22'}  S^\Lambda_{22'}(\Omega)\times\\
    &\Big[\bar v_{1'2'12}+ \gamma^{\mathrm{p},\Lambda}_{1'2'12}(\Omega+\omega)+\gamma^{\mathrm{x},\Lambda}_{1'2'12}(\Omega-\omega)+\gamma^{\mathrm{d},\Lambda}_{1'2'12}(0)\Big].
\end{split}\end{equation}
The channel decomposition makes the vertex flow equations manageable by numerics (each term $\gamma^{\alpha,\Lambda}$ depends only on a single frequency) but still include all terms of \(\ord{U^2}\).

In addition to decoupling the frequency structure, the channel decomposition also simplifies the dependence on the spatial indices of the vertex functions.  Since \(\gamma^{\alpha,\Lambda\to\infty}=0\), one trivially finds that in this limit:
\begin{equation}
    \label{eq:spatial_support}
    \begin{split}
        \gamma^{\mathrm{p},\Lambda\to\infty}_{1'2'12}(\Pi)&=0\hspace{0.5cm} \forall ~|1'-2'|\geq R ~\hspace*{-0.1cm} \lor~ |1-2|\geq R,\\
        \gamma^{\mathrm{x},\Lambda\to\infty}_{1'2'12}(X)&=0\hspace{0.5cm}  \forall ~|1'-2|\geq R ~\lor~ |2'-1| \geq R,\\
        \gamma^{\mathrm{d},\Lambda\to\infty}_{1'2'12}(\Delta)&=0\hspace{0.5cm}  \forall ~|1'-1|\geq R ~\lor~ |2'-2| \geq R,
    \end{split}
\end{equation}
where $|1-2|:=|i_1-i_2|$ refers to the distance of the single particle-indices within the multi-indices $1,2=(i_{1,2},\alpha_{1,2})$. Per the assumption in Eq.~(\ref{eq:support}), the same holds true for the single-particle structure of the initial vertex $\bar v$. One can easily see that the flow equations (\ref{eq:chan_decomp_flow}) preserve Eq.~(\ref{eq:spatial_support}), which thus remains true throughout the flow; e.g., the indices $1'$ and $2$ ($2'$ and $1$) appear as the first and last (second and third) argument on the rhs of the flow equation for $\gamma^{\mathrm{x},\Lambda}_{1'2'12}$. We emphasize that Eq.~(\ref{eq:spatial_support}) is a direct consequence of the channel decomposition and does not constitute an additional approximation.

\subsection{Making use of the system's symmetry}
The symmetries in Eq.~(\ref{eq:shift_sym_self_gf}) are self-consistently preserved within our approximation scheme (i.e., after truncation). If we assume that Eq.~(\ref{eq:shift_sym_self_gf}) holds for a given $\Sigma^\Lambda$ (and thus also for $G^\Lambda$ as well as $S^\Lambda$) and exploit that the initial, frequency-independent vertex fulfills Eq.~(\ref{eq:shift_sym}), we can use the flow equation (\ref{eq:chan_decomp_flow}) to show that
\begin{equation}\label{eq:shift_sym_vertex}
    \begin{split}
        \gamma^{\mathrm{p},\Lambda}_{(1'+L)(2'+L)(1+L)(2+L)}(\Pi)& = \gamma^{\mathrm{p},\Lambda}_{1'2'12}(\Pi-2LE),\\
        \gamma^{\mathrm{x},\Lambda}_{(1'+L)(2'+L)(1+L)(2+L)}(X)& = \gamma^{\mathrm{x},\Lambda}_{1'2'12}(X),\\
        \gamma^{\mathrm{d},\Lambda}_{(1'+L)(2'+L)(1+L)(2+L)}(\Delta) & = \gamma^{\mathrm{d},\Lambda}_{1'2'12}(\Delta).
    \end{split}
\end{equation}
If we now plug Eq.~(\ref{eq:shift_sym_vertex}) into the self-energy flow equation (\ref{eq:ford2}), it follows immediately that the symmetry relations in Eq.~(\ref{eq:shift_sym_self_gf}) are preserved.

In a nutshell, Eqs.~(\ref{eq:shift_sym_self_gf}) and (\ref{eq:shift_sym_vertex}) imply that one of the spatial indices (say \(i_{1'}\)) of the self-energy as well as of the two-particle vertex can be restricted to \(\{0,\dots,L-1\}\) when solving the flow equations.

\subsection{Integrations as convolutions}
\label{sec:frg_conv}

The flow equations (\ref{eq:chan_decomp_flow}) and (\ref{eq:ford2}) can all be rewritten in terms of convolutions:
\begin{equation}
    (f* g)(y)=\int \dOp x f(x)g(y-x).
\end{equation}
If we define the shorthand notation
\begin{equation}
    \tilde f(x)=f(-x),
\end{equation}
and split up the self-energy flow equation (\ref{eq:ford2}) into three terms, \(\Sigma^\Lambda=\sum_{\alpha\in\{p,x,d\}} \Sigma^{\alpha,\Lambda}\), we find
\begin{equation}\label{eq:flow_convo_fo}
    \begin{split}
        \partial_\Lambda \Sigma^{\mathrm{p},\Lambda}_{1'1}(\omega)&=-\frac{\I}{2\pi}\sum_{22'}\tilde S^\Lambda_{22'}*\gamma^{\mathrm{p},\Lambda}_{1'2'12},\\
        \partial_\Lambda \Sigma^{\mathrm{x},\Lambda}_{1'1}(-\omega)&=-\frac{\I}{2\pi}\sum_{22'}\tilde S^\Lambda_{22'}*\gamma^{\mathrm{x},\Lambda}_{1'2'12},\\
        \partial_\Lambda \Sigma^{\mathrm{d},\Lambda}_{1'1}(\omega)&=-\frac{\I}{2\pi}\sum_{22'}\bar \gamma^{\mathrm{d},\Lambda}_{1'2'12}(0)\int\dOp \Omega\, S^\Lambda_{22'}(\Omega).
    \end{split}
\end{equation}
No frequency-dependence is generated in the last term. Similarly, the flow equations (\ref{eq:chan_decomp_flow}) for the vertex can be recast as 
\begin{widetext}
\begin{equation}
    \label{eq:flow_convo}
    \begin{split}
        \partial_\Lambda \gamma^{\mathrm{p},\Lambda}_{1'2'12}(\Pi)&=\frac{\I}{2\pi}\sum_{33'44'}
                                                       \bar\gamma^{\mathrm{p},\Lambda}_{1'2'34}\left(\Pi\right)\left[G^\Lambda_{44'}*S^\Lambda_{33'}\right](\Pi)~\bar\gamma^{\mathrm{p},\Lambda}_{3'4'12}\left(\Pi\right),\\
        \partial_\Lambda \gamma^{\mathrm{x},\Lambda}_{1'2'12}(X)&=\frac{\I}{2\pi}\sum_{33'44'}
        \bar\gamma^{\mathrm{x},\Lambda}_{1'4'32}\left( X\right)\biggl[G^\Lambda_{44'}*\tilde S^\Lambda_{33'} +S^\Lambda_{44'}*\tilde G^\Lambda_{33'}\biggr](X)~\bar\gamma^{\mathrm{x},\Lambda}_{3'2'14}\left(X\right),\\
        \partial_\Lambda \gamma^{\mathrm{d},\Lambda}_{1'2'12}(\Delta)&=\frac{-\I}{2\pi}\sum_{33'44'}\bar\gamma^{\mathrm{d},\Lambda}_{1'3'14}\left(\Delta\right)\biggl[G^\Lambda_{44'}*\tilde S^\Lambda_{33'}+ S^\Lambda_{44'}*\tilde G^\Lambda_{33'}\biggr](\Delta)~\bar\gamma^{\mathrm{d},\Lambda}_{4'2'32}\left(\Delta\right).
    \end{split}
\end{equation}
\end{widetext}
This shows that a numerically-efficient implementation of the flow equations can be based on an efficient implementation of convolutions, which in turn can be achieved by employing fast Fourier transforms to perform all integrations (see Appendix~\ref{sec:convo}).

While at \(T=0\) some components of the Green's functions and single-scale propagators are discontinuous, this is not true for the vertex functions, which one can understand as follows: The rhs of the flow equations (\ref{eq:flow_convo}) is governed by a convolution of two functions $G^\Lambda$ and $S^\Lambda$ that decay sufficiently quickly for \(\omega\to\pm\infty\); this yields a continuous function.

\subsection{Support of the vertex functions}
\label{sec:frg_support}

During the flow, self-energy components with arbitrary single-particle indices are generated by Eq.~(\ref{eq:flow_convo_fo}). The same holds true for the two-particle vertex with the exception that the spatial structure of Eq.~(\ref{eq:spatial_support}) is always preserved (e.g., components with arbitrary $i_{1'}-i_1$ can be generated in $\gamma^{\mathrm{p},\Lambda}_{1'2'12}$). Moreover, the rhs of Eqs.~(\ref{eq:flow_convo_fo}) and (\ref{eq:flow_convo}) contains infinite sums over single-particle indices. Thus, we need to devise additional approximations in order to make a numerical treatment feasible. To this end, we introduce \( M\in\mathbb{N},M\geq R\) as a cutoff parameter and set
\begin{alignat}{3}
    \label{eq:truncM}
    \Sigma_{1'1}^\Lambda&=0\hspace{2mm}&&\forall &|1-1'|&\geq M, \\
    \gamma_{1'2'12}^\Lambda&=0\ &&\forall\ &\text{dist}(1',2',1,2)&\geq M,\notag
\end{alignat}
where $|1-1'|:=|i_1-i_{1'}|$ again refers to the distance of the single-particle indices, and the more special case of Eq.~(\ref{eq:spatial_support}) always holds exactly. This is a natural assumption in a system where inelastic scattering limits the correlation length. 
In the limit \(M\to\infty\), we recover the original flow equations (\ref{eq:chan_decomp_flow}) and (\ref{eq:ford2}). Note that this choice of truncation preserves the symmetries\cite{Jakobs2010}
\begin{equation}
    \begin{split}
        \Sigma^{\Lambda}_{1'1}(\omega_{1'}, \omega_1)=&(-1)^{\alpha_1+\alpha_{1'}}\left[\Sigma^{\Lambda}_{\bar{1} \bar{1'}} (\omega_1, \omega_{1'})\right]^*,\\
        \gamma^\Lambda_{1'2'12}(\omega_{1'},\omega_{2'},\omega_1, \omega_2)=&-\gamma^\Lambda_{2'1'12}(\omega_{2'},\omega_{1'},\omega_1, \omega_2),\\
        =&-\gamma^\Lambda_{1'2'21}(\omega_{1'},\omega_{2'},\omega_2, \omega_1),\\
        \gamma^\Lambda_{1'2'12}(\omega_{1'},\omega_{2'},\omega_1, \omega_2)=&(-1)^{1+\alpha_1+\alpha_2+\alpha_{1'}+\alpha_{2'}}\times\\
                                                                            &\times\gamma^\Lambda_{\bar 1 \bar 2 \bar{1'}\bar{2'}}(\omega_1, \omega_2,\omega_{1'},\omega_{2'})^*,\\
    \end{split}
\end{equation}
where \(\bar{1}=(i_1, 3-\alpha_1)\). This is essential in order to preserve the fluctuation-dissipation theorem in the equilibrium limit [see Eq.~(\ref{eq:fluc_dis})].

Overall, we are left with a maximum of \(\sim L MR^2\) independent, frequency-dependent components of the two-particle vertex functions. Note that Eq.~(\ref{eq:truncM}) implies that the summations on the lhs of Eqs.~(\ref{eq:flow_convo_fo}) and (\ref{eq:flow_convo}) are limited by $\tn{dist}(2,2')< M$ and $\tn{dist}(3,3',4,4')< 3M$, respectively; it is thus sufficient to set $N=3M$ when calculating $G^\Lambda$ and $S^\Lambda$ (see Sec.~\ref{sec:iterative_gf}).

\subsection{Single-scale propagators}\label{sec:iterative_s}

In Sec.~\ref{sec:iterative_gf}, we discussed how the Green's functions $G^\Lambda$ of an infinite system can be computed iteratively. Since the rhs of the FRG flow equations also contains the single-scale propagator $S^\Lambda$, we will now illustrate how this quantity can be computed along the lines of Sec.~\ref{sec:iterative_gf}. To improve readability, we will frequently refrain from writing out the $\Lambda$-dependence as well as frequency-arguments of various quantities such as $G^\Lambda(\omega)$ and $S^\Lambda(\omega)$ throughout this section.

\subsubsection{Retarded single-scale propagator}
We begin with the retarded part of the single-scale propagator; the advanced part follows from $[S^\tn{ret}]^\dagger = S^\tn{adv}$. Since our cutoff stems from wide-band reservoirs coupled to each site, it only enters on the diagonal of $[G^\tn{ret}]^{-1}$ and thus
\begin{equation}
    \begin{split}
    \pa D = \I,~~
        \partial_\Lambda^* T_{\ci\li}=\partial_\Lambda^* T_{\li\ci}=\partial_\Lambda^* T_{\ci\ri}=\partial_\Lambda^* T_{\ri\ci}=0.
    \end{split}
\end{equation}
This restriction reduces the number of terms in the following expressions, but a generalization to a more involved cutoff scheme is straightforward.

The retarded part of the single-scale propagator can be computed by taking the derivative of Eq.~(\ref{eq:iter_greta}):
\begin{equation}
\begin{split}
 S^\tn{ret}_{\ci\ci} = \partial_\Lambda^* G^\tn{ret}_{\ci\ci} = - G^\tn{ret}_{\ci\ci}\Big[ 
\I  &- T_{\ci\li} \calSrLR_{\li\li}T_{\li\ci} \\
& - T_{\ci\ri} \calSrLR_{\ri\ri}T_{\ri\ci}
\Big] G^\tn{ret}_{\ci\ci}.
\end{split}
\end{equation}
The only unknown quantity in this equation is $\calSrLR:=\pa\calrLR$, which can be obtained via the derivative of Eq.~(\ref{eq:iter_gf2}):
\begin{equation}\label{eq:iter_sr2}
\begin{split}
&~ \calSrLR_{\li\li}(\omega-NE)\\  =&-\calrR_{\ci\ci}(\omega)\Big[\I - T_{\ci\li}\calSrLR_{\li\li}(\omega)T_{\li\ci}\Big]\calrR_{\ci\ci}(\omega),
\end{split}
\end{equation}
and similarly for $\calSrLR_{\ri\ri}$. Yet again, this equation can either be solved self-consistently (at \(E=0\)) or successively (for \(E\neq0\)) using the boundary conditions \(\calSrLR(\pm\infty)=0\) as outlined in Appendix~\ref{sec:sol_self_cons}.

\subsubsection{Keldysh single-scale propagator}

We now proceed with the Keldysh component of the single-scale propagator. The cutoff only enters into the diagonal of $\tsk$, which leads to [see Eqs.~(\ref{eq:self_hybrid}), (\ref{eq:tsk}), and (\ref{eq:frgcutoff})]
\begin{equation}
    \begin{split}
        \partial_\Lambda^* \tsk_{ij}(\omega)& =\partial_\Lambda \sum_\nu \Gamma^{\nu,\mathrm{K}}_{ij}(\omega)\\
        &=2\I\sum_{\nu} [1-2n^\nu(\omega)]\partial_\Lambda \Im \Gamma^{\nu,\mathrm{ret}}_{ij}(\omega) \\
        & =2\I[1-2n^i(\omega)]\delta_{ij}.
    \end{split}
\end{equation}
Taking the derivative of Eq.~(\ref{eq:iter_gk1}) yields (all quantities carry a frequency argument $\omega$)
\begin{equation}
\begin{split}
 S^\tn{K}_{\ci\ci}(\omega) = S^\tn{ret}_{\ci\ci}(\omega)\Big[&\cdots\Big]G^\tn{adv}_{\ci\ci}(\omega)+G^\tn{ret}_{\ci\ci}(\omega)\Big[\cdots\Big]S^\tn{adv}_{\ci\ci}(\omega) \\
+ G^\tn{ret}_{\ci\ci}\Big[ \pa\tsk_{\ci\ci} - & T_{\ci\li}\calSrLR_{\li\li}\tsk_{\li\ci} - \tsk_{\ci\li} \calSaLR_{\li\li} T^\dagger_{\li\ci} \\
 -&T_{\ci\ri}\calSrLR_{\ri\ri}\tsk_{\ri\ci}  - \tsk_{\ci\ri} \calSaLR_{\ri\ri} T^\dagger_{\ri\ci} \\
  +& T_{\ci\li}\calSkLR_{\li\li} T^\dagger_{\li\ci}+ T_{\ci\ri}\calSkLR_{\ri\ri}T^\dagger_{\ri\ci}\Big]G^\tn{adv}_{\ci\ci},
\end{split}
\end{equation}
where the brackets $[\ldots]$ are identical to the bracket in Eq.~(\ref{eq:iter_gk1}). The only unknown quantity is $\calSkLR_{\li\li}:=\pa\calkLR_{\li\li}$, which can be determined by taking the derivative of Eq.~(\ref{eq:iter_gk2}):
\begin{equation}\label{eq:iter_sk2}
\begin{split}
&~ \calSkLR_{\li\li}(\omega-NE) \\= &~\calSrR_{\ci\ci}(\omega)\Big[\cdots\Big]\calaR_{\ci\ci}(\omega)+\calrR_{\ci\ci}(\omega)\Big[\cdots\Big]\calSaR_{\ci\ci}(\omega) \\
+ &~ \calrR_{\ci\ci}\Big[ \pa\tsk_{\ci\ci} -  T_{\ci\li}\calSrLR_{\li\li}\tsk_{\li\ci} - \tsk_{\ci\li} \calSaLR_{\li\li} T^\dagger_{\li\ci} \\
  &\hspace*{1.2cm}+ T_{\ci\li}\calSkLR_{\li\li} T^\dagger_{\li\ci}\Big]\calaR_{\ci\ci},
\end{split}
\end{equation}
where the brackets $[\ldots]$ are identical to the bracket in Eq.~(\ref{eq:iter_gk2}), and we have omitted frequency arguments $\omega$ in the last two lines. Eq.~(\ref{eq:iter_translation2}) implies that $\calSrR_{\ci\ci}(\omega)=\calSrLR_{\li\li}(\omega-NE)$, which was calculated via Eq.~(\ref{eq:iter_sr2}). Eq.~(\ref{eq:iter_sk2}) can be solved self-consistently ($E=0$) or successively ($E\neq0$, see Appendix~\ref{sec:sol_self_cons}).

\subsection{Frequency discretization}
\label{sec:freq_disc}
For a numerical treatment, it is necessary to discretize the frequency space. In order to faithfully represent the physical system at hand, one must choose a grid that accounts for all of its relevant energy scales. For simplicity, we evaluate both the vertex functions as well as the Green's functions and single-scale propagators on the same set of frequencies.

The Green's functions decay on the scale of the system's bandwidth and are broadened by inelastic scattering.
Furthermore, the Keldysh Green's function is linked to the distribution function within the reservoirs via Eq.~\eqref{eq:gk_from_self}, making the temperature of the reservoirs a relevant energy scale. This motivates the use of an equidistant grid whose width scales with the bandwidth and which includes additional points around the chemical potentials of the reservoirs. For the case of zero-temperature reservoirs, it is most convenient to choose
\begin{equation}
    \begin{split}
        \Omega&=\Omega_\mathrm{equi}\cup\Omega_\mathrm{extra},\\
        \Omega_\mathrm{equi}&=\left\{n\delta_\omega \middle| n\in \mathbb{Z},\ |n\delta_\omega|\leq \omega_\mathrm{max} \right\},\\
        \Omega_\mathrm{extra}&=\bigcup\limits_{\nu=-N^\mu_\mathrm{max}}^{N^\mu_\mathrm{max}}\left\{\mu_\nu-\epsilon,\mu_\nu+\epsilon \right\},
    \end{split}
\end{equation}
with \(\epsilon\ll \delta_\omega\). At non-zero temperature, a more sophisticated choice of \(\Omega_\mathrm{extra}\) is required.
In Appendix~\ref{sec:convo}, we show that using such a grid allows for an efficient implementation of convolutions using fast Fourier transforms if the number of points in $\Omega_\tn{extra}$ is small.

When the coupling to the reservoir dominates all other energy scales, we require that \(\omega_\mathrm{max}\gg \Lambda \gg \delta_\omega\). In the opposite limit, the bandwidth \(B=4t\) of the closed system determines the width of the grid, \(\omega_\mathrm{max}\gg B\), and \(\delta_\omega^{-1}\) should be chosen much larger than the coherence time of the closed system.
If not otherwise stated, we use
\begin{equation}
    \begin{split}
    \omega_\mathrm{max}&=20\Lambda + 16t,\\
    2\omega_\mathrm{max}/\delta_\omega&=3^7,\\
    \epsilon&=10^{-4}t,\\
    N^\mu_\mathrm{max}&=12.
    \end{split}
\end{equation}
Since \(\omega_\mathrm{max}\) depends on $\Lambda$, it is necessary to adapt the grid during the solution of the flow equations. This is done after every step of the differential equation solver; in order to obtain the vertex functions on the new grid, we use linear interpolation (note that extrapolation is never required as \(\omega_\mathrm{max}\) is only decreased).

\section{Application}
\label{sec:application}

We will now apply our iterative Green's function algorithm as well as our novel non-equilibrium FRG approach to the tight-binding chain introduced in Sec.~\ref{sec:class_of_models}. A pictorial representation of this model is shown in Fig.~\ref{fig:pic_chain}.

First, we will benchmark the iterative Green's function algorithm introduced in Sec.~\ref{sec:iterative_gf} against analytical results available for the non-interacting Hamiltonian $U=0$. In this limit, all single-particle eigenfunctions of the closed system ($\Gamma=0$) are exponentially localized for any non-zero $E$, leading to Wannier-Stark insulating behavior. The conductivity becomes finite for any $\Gamma>0$.

Secondly, we will explore the capabilities of our novel non-equilibrium functional RG approach in describing finite interactions $U>0$. There are two reference results that we can compare against. First, we will set up a mean-field treatment to compute the phase diagram in the large-$U$ limit; this serves as a highly non-trivial test for the FRG, which is perturbative w.r.t.~$U$. Secondly, it is known that the closed system with zero electric  field (\(\Gamma=E=0\)) undergoes a Berezinskii-Kosterlitz-Thouless quantum phase transition from a gapless Tomonaga-Luttinger liquid (\(U<2t\)) to a charge density wave (\(U>2t\)). The order parameter of the charge density wave (CDW) phase can be defined as the occupation difference between even and odd sites:
\begin{equation}\label{eq:def_deln}
\Delta n = \langle \hat n_\tn{even}\rangle-\langle \hat n_\tn{odd}\rangle.
\end{equation}
The corresponding susceptibility
\begin{equation}
\chi = \lim_{s\to0}\frac{\Delta n}{s}
\end{equation}
diverges in a CDW phase but is finite otherwise. We will explicitly compute these quantities using the FRG for arbitrary $U$, $\Gamma$, and $E$.

\begin{figure}
    \begin{overpic}[width=0.47\columnwidth]{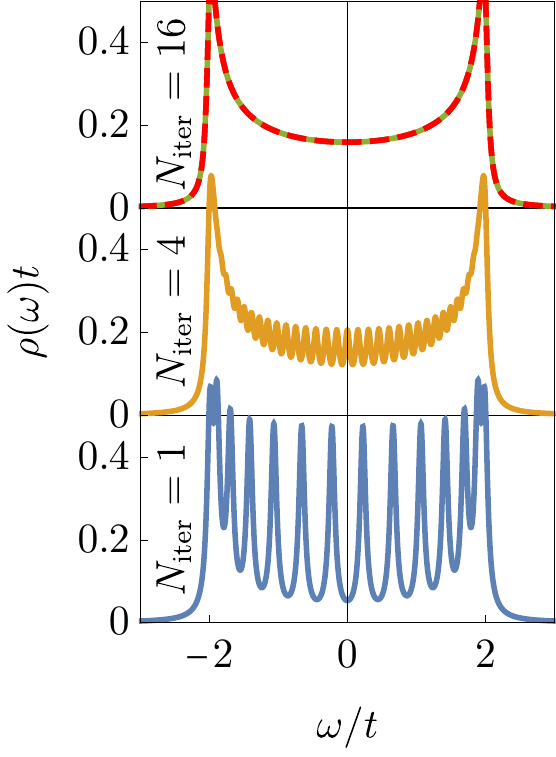}
        \put(0,95){(a)}
    \end{overpic}\hspace*{0.03\linewidth}
    \begin{overpic}[width=0.47\columnwidth]{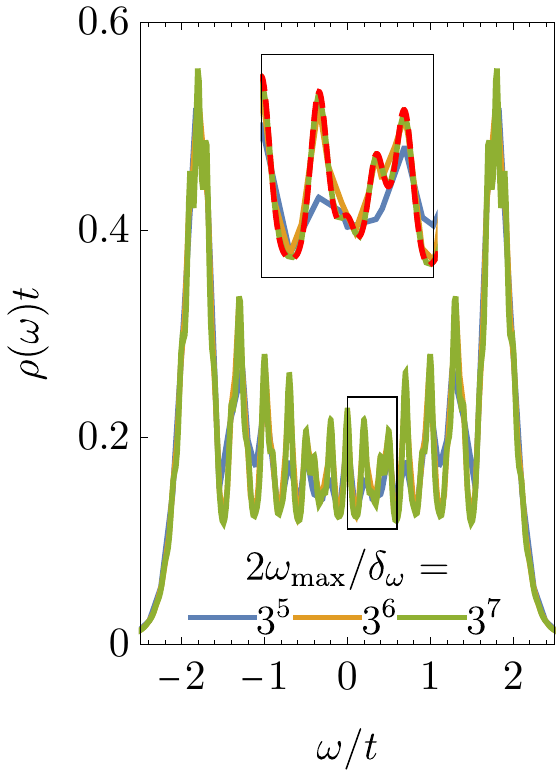}
        \put(0,95){(b)}
    \end{overpic}
    \caption{
        Benchmarking of the iterative, infinite-system Green's function algorithm introduced in Sec.~\ref{sec:iterative_gf} in the non-interacting limit $U=0$.
        (a) Local density of states \(\rho(\omega)=-\frac{1}{\pi}\Im[g^\mathrm{ret}(\omega)]\)  at \(\Gamma/t=0.05\) and \(E=0\) with \(1,\ 4,\ 16\) iterations of the self-consistency equation~\eqref{eq:iter_gf2} (blue, yellow and green; the curves are shifted vertically for readability ).
        The dashed red line indicates the analytic result of Eq.~(\ref{eq:dosexact}).
        (b) The same but for a finite electric field \(E/t=0.1\). Results are obtained using a frequency grid with \(2\omega_\mathrm{max}/\delta_\omega=3^\alpha,\ \alpha=5,6,7\), which corresponds to a grid spacing of \(\delta_\omega/t=0.14,\ 0.047,\ 0.016\). 
        The inset shows a zoom-in of the region indicated in the main panel, and the dashed red line shows the analytical solution for a system of \(61\) sites.
    }\label{fig:dos}
\end{figure}

\subsection{Non-interacting properties}\label{ssec:noninterprops}

\subsubsection{Density of states}

We now discuss the physics for $U=0$ in more detail and use this limit as a testing ground for our iterative Green's function algorithm. In the absence of an electric field, the local density of states (LDOS) is known analytically:
\begin{equation}\label{eq:dosexact}
    \rho(\omega)=-\frac{1}{\pi}\Im\left(\frac{1}{\sqrt{\left(\omega+\I\Gamma\right)^2-4t^2}}\right).
\end{equation}
In Fig.~\ref{fig:dos}(a), we demonstrate how this exact result is recovered by the iterative algorithm of Sec.~\ref{sec:iterative_gf}.

For finite fields \(E>0\) and $\Gamma=0$, the system is a Wannier-Stark insulator, and all single-particle eigenstates are exponentially localized.
At small \(\Gamma\ll t\), the local density of states becomes sharply peaked on the scale of 
\(\Gamma\). In Fig.~\ref{fig:dos}(b), we demonstrate how our iterative algorithm can be used to compute the LDOS in a numerically-exact fashion. (One should note that for finite $U$, additional inelastic processes lead to smoother Green's functions, which simplifies computations.)

\subsubsection{Current}

Wannier-Stark localization has a profound impact on the current $I$ flowing through the system in the presence of a finite electric field.
In the absence of interactions, the current through a bond, say the one connecting site \(0\) to site \(1\), is given by\cite{Datta1995}
\begin{equation}\label{eq:currentdef}
    \begin{split}
        I&=\sum_{\nu\leq0}\sum_{\nu'\geq 1} I_{\nu\nu'}, \\
        I_{\nu\nu'}&=\frac{2}{\pi}\int \dOp \omega \left[ n^\nu(\omega)-n^{\nu'}(\omega) \right]\Gamma^2 \left|g^\mathrm{ret}_{\nu\nu'}(\omega)\right|^2,
\end{split}
\end{equation}
where \(I_{\nu\nu'}\) can be understood as the current between reservoirs \(\nu\) and \(\nu'\). One can show that for small fields $E\ll t$, the hybridization-dependence $I(\Gamma)$ reads (note that the bandwidth is given by $B=4t$):\footnote{The current in the absence of interactions was computed in Ref.~\onlinecite{Han2013}, and analytical formulas for $I(\Gamma)$ were derived in the limit $E,\Gamma\ll t$. We briefly recapitulate these results for didatic purposes.}
\begin{equation}\label{eq:currentgamma}
    I\sim \begin{cases}
            \frac{\Gamma t}{E} & \Gamma\ll E \\
            \frac{Et}{\Gamma} & E\ll \Gamma \ll t\\
            \frac{Et^2}{\Gamma^2} &t \ll \Gamma .
    \end{cases}
\end{equation}
The behavior in these three regimes can be understood qualitatively.
For small $\Gamma$, Wannier-Stark localization leads to a vanishing current; the number of particles entering the chain from the reservoirs scales with \(\Gamma\). Once in the system, each fermion is repeatedly reflected by the electric field in the form of Bloch oscillations. The typical distance traveled before eventually leaving the chain scales as \(t/E\), resulting in a total current that scales as \(\Gamma t/E\).

\begin{figure}
    \includegraphics[width=\columnwidth]{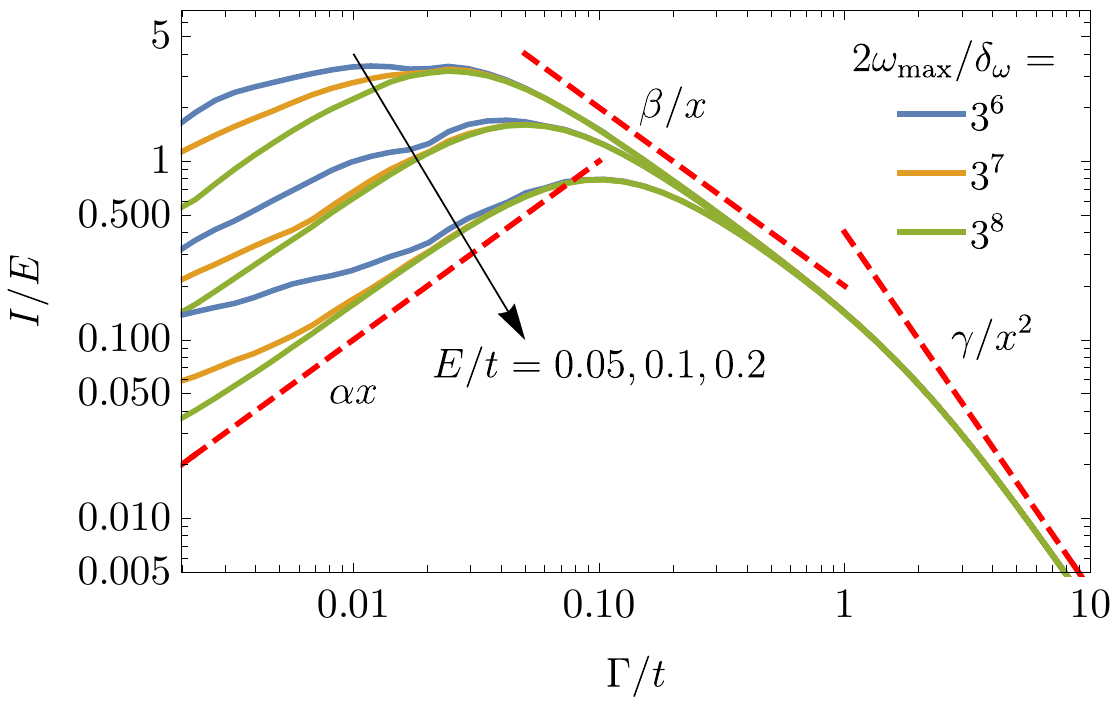}
    \caption{ The finite-field conductivity \(I/E\) as a function of the reservoir coupling \(\Gamma\) for three different fields $E$ in absence of interactions ($U=0$). According to Eq.~(\ref{eq:currentgamma}), one expects three distinct power-law regimes. The data was computed using the iterative algorithm of Sec.~\ref{sec:iterative_gf}.  The different colors indicate different spacings  $\delta_\omega$ of the frequency grid, indicating that convergence can be reached.
    }\label{fig:non_int_cond}
\end{figure}

If \(E \ll \Gamma\lesssim t\), fermions tunnel coherently between far-apart reservoirs.
To analyze this in more detail, consider the defining equation for an element of the non-interacting Green's function:
\begin{equation}
    \sum_k \left[(\omega-kE+\I\Gamma)\delta_{i,k}-t\delta_{|i-k|,1}\right] g_{kj}(\omega)=\delta_{i,j}.
\end{equation}
At \(\omega=0\), \(E\ll\Gamma\lesssim t\), and \(i\ll j\) or \(j\ll i\), this equation is approximately solved by 
\begin{equation}
    g_{ij}(\omega=0)\sim-\I^{|i-j|}\exp(-|i-j|\Gamma/2t),
\end{equation}
only resulting in deviations of \(\mathcal{O}(\Gamma^3)\). We plug this into Eq.~(\ref{eq:currentdef}) and exploit that for a given distance \(\Delta\) between pairs of reservoirs, there is \(\mathcal{O}(\Delta)\) pairs that are connected by the selected bond. The function \(n^\nu(\omega)-n^{\nu'}(\omega)\) has a width \(\Delta E\), and the Green's function can be assumed to be constant within such an interval. Therefore, the total current through that bond is obtained as
\begin{equation}
    \label{eq:loc_cur}
    I\sim\sum_{\Delta=1}^\infty \frac{E \Delta^2\Gamma^2}{t^2} \exp(-\Delta\Gamma/t)\sim \frac{Et}{\Gamma}.
\end{equation}
At large coupling \(\Gamma\gg t\), the scattering into the reservoirs dominates and the correlation becomes small [compare Eq.~(\ref{eq:dysonret})]:
\begin{equation}
    \begin{split}
    & \gr(\omega) =\frac{1}{\omega-h+\I\Gamma}=\frac{1}{\I\Gamma}+\frac{\omega-h}{\Gamma^2}+\mathcal{O}\left(\frac{1}{\Gamma^3}\right),\\
    & \gr_{i(i+1)}(\omega=0) \sim\frac{t}{\Gamma^2}\ \Rightarrow\ I\sim\frac{Et^2}{\Gamma^2}.
\end{split}
\end{equation}
The current obtained using our numerical algorithm is shown in Fig.~\ref{fig:non_int_cond} as a function of $\Gamma$ for various values of $E<t$. The three distinct regimes of Eq.~(\ref{eq:currentgamma}) can be clearly identified; this serves as an additional benchmark for our iterative approach.

\subsection{Large interaction limit}\label{ssec:large_inter}

In the previous section, we have tested the (numerically exact) iterative algorithm introduced in Sec.~\ref{sec:iterative_gf} by comparing with analytical results in the non-interacting limit $U=0$. We now turn to benchmarking our FRG approximation scheme for $U\neq0$. To the best of our knowledge, no reference data is available except in the limit $E=\Gamma=0$. We therefore resort to a mean-field treatment, which is expected to give reasonable results for $U\to\infty$. This is a highly non-trivial testing ground for the FRG, which is perturbative w.r.t.~the interaction strength.

\subsubsection{Mean-field approach}

If the interaction dominates the bare, decoupled Hamiltonian (i.e., for \(U\gg t\)), the system can be reduced to a chain of disconnected sites. One can easily show that for $t=0$, the effects of a finite field \(E\neq0\) can be eliminated by virtue of a gauge transformation analogous to the Peierls substitution:\cite{Peierls1933}
\begin{equation}
    \begin{split}
        c_j &\to \exp(-\I jE\tau)c_j,\\
        a_{k,j} &\to \exp(-\I jE\tau) a_{k,j},
    \end{split}
\end{equation}
where \(\tau\) denotes time (within the action). This transformation effectively shifts all energies on site \(j\) and the adjacent reservoir by \(-jE\).
Hence, we can restrict ourselves to \(E=0\). For $\Gamma=0$, the ground state is a perfect charge density wave, while for large couplings $\Gamma\gg U$, the ground state is not spontaneously ordered. To our best knowledge, the critical coupling strength which characterizes the transition between these two regimes is not known. 

It is reasonable to treat the limit $t=0$ using mean-field theory. Since the Green's functions become diagonal, one needs to solve the following self-consistency equation:
\begin{equation}
  \begin{split}
      \left\langle  \hat n_\mathrm{even}\right\rangle=&\frac{-1}{\pi}\int_{-\infty}^{0}\mathrm{d}\omega\, \Im\left(\frac{1}{\omega+s+U-2U\left\langle \hat n_\mathrm{odd} \right\rangle+\I \Gamma}\right).
  \end{split}
\end{equation}
Using Eq.~\eqref{eq:def_deln} as well as \(\left\langle
  \hat n_\mathrm{even}\right\rangle=1-\left\langle \hat
  n_\mathrm{odd}\right\rangle\), this reduces to
\begin{equation}
  \Delta n=\frac{2}{\pi}\arctan\left( s/\Gamma+U\Delta n/\Gamma \right).
  \end{equation}
At \(s=0\), this equation has nontrivial solutions beyond a critical interaction strength of
\begin{equation}\label{eq:ucritmf}
    U^\mathrm{MF}_\mathrm{crit}=\frac{\pi}{2}\Gamma^\mathrm{MF}_\mathrm{crit}\approx 1.571\Gamma^\mathrm{MF}_\mathrm{crit}.
\end{equation}
In the disordered phase (\(U/\Gamma<\pi/2\)), the susceptibility for small
symmetry breaking (\(s\ll U,\Gamma\)) scales as
\begin{equation}
  \chi^\mathrm{MF}=\lim_{s\to 0}\frac{\Delta n}{s}\sim \frac{1/\Gamma}{\pi/2-U/\Gamma},\hspace{0.5cm} U/\Gamma<\pi/2.
  \end{equation}

For reasons of completeness, we will now present technical details of how to solve the mean-field equations for $t\neq0$; results will be discussed elsewhere. In thermal equilibrium, a common technique to finding solutions of the mean-field equations is to use a self-consistency loop based on an initial guess for the
field. If multiple solutions are found, one picks the one which minimizes the free energy; this solution corresponds to a stable fixed point of the mean-field equations.

For finite fields $E\neq 0$, the above procedure needs to be modified. First, a self-consistency loop is insufficient to identify all solutions of the mean-field equations as not all fixed points can be obtained as the limit of a self-consistency loop, regardless of the initial guess. One therefore needs to resort to a version of Newton's method, which also converges to fixed points where a self-consistency loop fails. Secondly, the free energy can no longer be used to determine a unique, stable solution. One can gauge the stability from the Lipschitz constant, which is estimated from the behavior of the self-consistency loop under a small perturbation. Out of equilibrium, one can thus only report the existence of solutions of the self-consistent equations and their stability with respect to perturbation. This will be discussed in a separate publication.

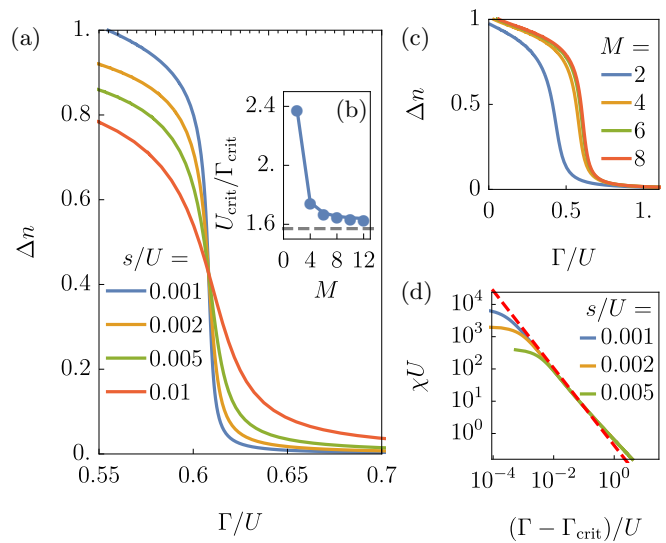
\begin{figure}
    \begin{overpic}[width=\columnwidth]{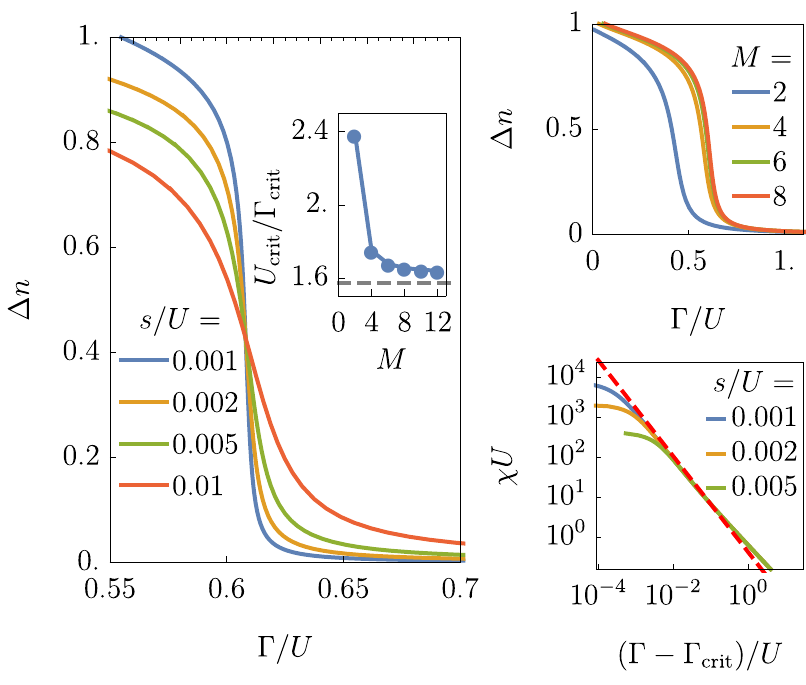}
        \put(60,78){(c)}
        \put(60,39){(d)}
        \put(0,78){(a)}
        \put(50,67){(b)}
    \end{overpic}
    \caption{Functional RG results in the limit $U\gg t$.
        (a) Difference $\Delta n$ between the occupation of even and odd sites (the CDW order parameter) as a function of the reservoir coupling $\Gamma$ for various values of the initial symmetry breaking $s$. The transition into the CDW phase becomes sharper as $s$ is decreased. The data was obtained for fixed $M=10$, which is the numerical control parameter in the solution of the flow equations.
        (b) \(M\)-dependence of the critical ratio \(U_\mathrm{crit}/\Gamma_\mathrm{crit}\), which is defined as the point in (a) where lines at different \(s\) cross. The dashed line shows the mean-field result of Eq.~(\ref{eq:ucritmf}).
        (c) CDW order parameter as a function of $\Gamma$ for fixed $s/U=0.01$ but different $M$, which again illustrates that convergence in $M$ can be reached.
        (d) The susceptibility $\chi=\lim_{s\to0}\Delta n / s$ features a power-law divergence with an exponent \(\gamma\approx 1.2\) (dashed line) close to the critical point.
    }\label{fig:large_inter_phase}
\end{figure}

\subsubsection{Benchmarking the FRG}
  
The mean-field results in the limit $U\to\infty$ can be used to benchmark our FRG approach (which is perturbative w.r.t.~$U$). Since $h$ is diagonal, the same holds true for all single-particle propagators $g$ (which can thus be computed straightforwardly without resorting to the iterative algorithm of Sec.~\ref{sec:iterative_gf}) as well as for the self-energy.  This allows us to greatly simplify the FRG flow equations. Note that the two-particle vertex functions remain infinitely extended as terms with arbitrary $\textnormal{dist}(1',2')$ and $\textnormal{dist}(1,2)$ can be generated by the flow. However, \(\gamma^\Lambda_{1'212}=0\) if \(\mathrm{dist}(1',1)>0\), and the self-energy itself is strictly local.

The main FRG results are presented in Fig.~\ref{fig:large_inter_phase}(a). For large reservoir couplings $\Gamma$, the CDW order parameter $\Delta n$ scales linearly in the initial symmetry breaking \(s\), which corresponds to a finite susceptibility $\chi$. Below a critical coupling, however, $\Delta n$ takes a finite value which does not decrease with decreasing \(s\); the response of the system diverges, and translation invariance is spontaneously broken. One can identify the critical point $\Gamma_\mathrm{crit}$ of this transition as the point where the curves at different $s$ intersect. In Fig.~\ref{fig:large_inter_phase}(b), we show $\Gamma_\mathrm{crit}$ as a function of \(M\), which is the control parameter in the numerical solution of the FRG flow equations. For large $M$, we obtain
\begin{equation}\label{eq:ucrit}
    U_\mathrm{crit}\approx 1.639\Gamma_\mathrm{crit},
\end{equation}
which is in good agreement with the mean-field prediction of Eq.~(\ref{eq:ucritmf}). For reasons of completeness, we show the CDW order parameter for different values of $M$ in Fig.~\ref{fig:large_inter_phase}(c), which again confirms that convergence can be reached. At the critical point, the susceptibility appears to diverge as a power law, and the FRG result for the critical exponent reads:
\begin{equation}
    \chi\sim \alpha \left(\frac{\Gamma-\Gamma_\mathrm{crit}}{U}\right)^{-\gamma},\ \gamma\approx1.2.
\end{equation}
This is illustrated in Fig.~\ref{fig:large_inter_phase}(d).

\begin{figure}
    \begin{overpic}[width=1.0\columnwidth]{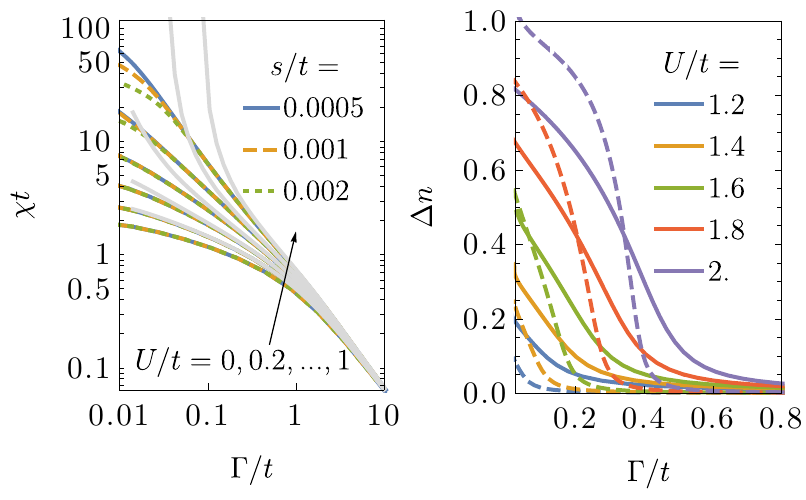}
        \put(0,60){(a)}
        \put(50,60){(b)}
    \end{overpic}
    \caption{
        (a) Functional RG results for the susceptibility $\chi$ for finite interactions $U/t\lesssim1$, $E=0$, and different values for the initial symmetry breaking $s$. The susceptibility is finite (though large) for any $\Gamma$. The FRG calculation was performed with \(M=4\). 
        In contrast to the FRG, mean-field theory (gray lines) predicts a transition into a CDW phase at any $U$ for small enough $\Gamma$.
        (b) For larger interactions and small $\Gamma$, the response of the system is no longer linear in \(s\), as demonstrated here for \(s/t=0.01\) (solid) and \(s/t=0.001\) (dashed). This indicates a divergent susceptibility and a transition into a CDW phase.
    }\label{fig:dis_phase}
\end{figure}

\begin{figure}
    \begin{overpic}[width=\columnwidth]{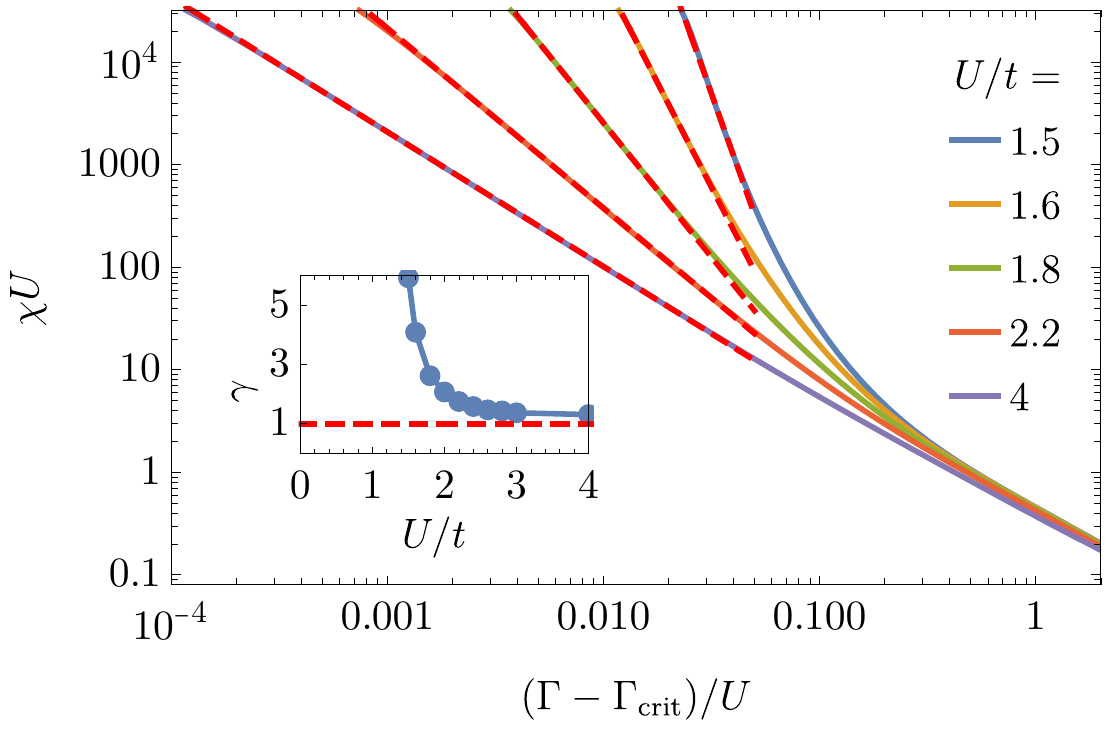}
        \put(2,64){(a)}
        \put(18,40){(b)}
    \end{overpic}
    \caption{
        FRG results for the critical behavior of the susceptibility at $E=0$ close to the phase transition. The data was obtained using \(s/U=10^{-6}\) and \(M=4\).
        (a) The susceptibility diverges as a power-law \(\chi\sim (\Gamma-\Gamma_\mathrm{crit})^{-\gamma}\) (dashed red lines).
        (b) The corresponding exponent $\gamma$ is interaction-dependent and diverges at a finite value $U_\tn{crit}$. The FRG does not predict a transition into a CDW phase at low interactions. In contrast, mean-field theory (dashed red line) yields a transition for arbitrary $U$ with an exponent \(\gamma^\mathrm{MF}=1\).
    }\label{fig:critical}
\end{figure}

\subsection{Phase diagram at intermediate interaction}

At intermediate interaction, the ordering tendencies driven by the interaction compete with the kinetic energy, resulting in a non-trivial phase-diagram.
In the absence of an electric field, mean field theory predicts an ordered phase at {\it any} interaction $U$ when the coupling \(\Gamma\) is small enough ($U_\mathrm{crit}/t\to0$ for $\Gamma\to 0$). In equilibrium and for $\Gamma\to 0$ this is known to be an artifact of the mean-field approximation. The exact Bethe-ansatz solution predicts that the CDW order is destabilized by quantum fluctuations and that a finite critical $U_\mathrm{crit}/t=2$ is needed to drive the phase into an ordered state by a Berezinskii-Kosterlitz-Thouless mechanism.\cite{DesCloizeaux1966,Giamarchi2006}

In contrast to the mean-field approach and in accord with the exact solution, we do not observe symmetry breaking at small interactions within the FRG calculation [see Fig.~\ref{fig:dis_phase}(a)]. While a phase transition at lower, inaccessible reservoir couplings $\Gamma$ can in principle not be ruled out, an alternative, Matsubara FRG scheme carried out directly at $\Gamma=0$ predicts a finite value of $U_\mathrm{crit}/t$.\cite{Markhof2018} Beyond a critical interaction of \(U_\mathrm{crit}/t\approx1.4\), the initial symmetry breaking yields a response that does not vanish linearly for $s\to0$, and the system enters a CDW phase [see Fig.~\ref{fig:dis_phase}(b)]. 

After pinpointing the position of the phase transition, we can analyze the critical behavior itself. Mean-field theory predicts a divergence as
\begin{equation}
    \chi^\mathrm{MF}\sim (\Gamma-\Gamma_\mathrm{crit}^\mathrm{MF})^{-\gamma^\mathrm{MF}},\ \gamma^\mathrm{MF}=1,
\end{equation}
independent of the interaction strength. In contrast, the susceptibility obtained using the FRG diverges as a power-law with an interaction-dependent exponent [see Fig.~\ref{fig:critical}(a)]. The exponent increases upon lowering the interaction towards the critical point, beyond which no phase transition can be identified at any $\Gamma>0$ [see Fig.~\ref{fig:critical}(b)]. The critical interaction strength extracted from the analysis is approximately $ U_\mathrm{crit}/t\approx 1.4 $ in the small-$\Gamma$ limit (the exact solution yields $U_\mathrm{crit}/t=2$).

\subsection{Phase diagram at non-zero electric field}
\begin{figure}
    \begin{overpic}[width=\columnwidth]{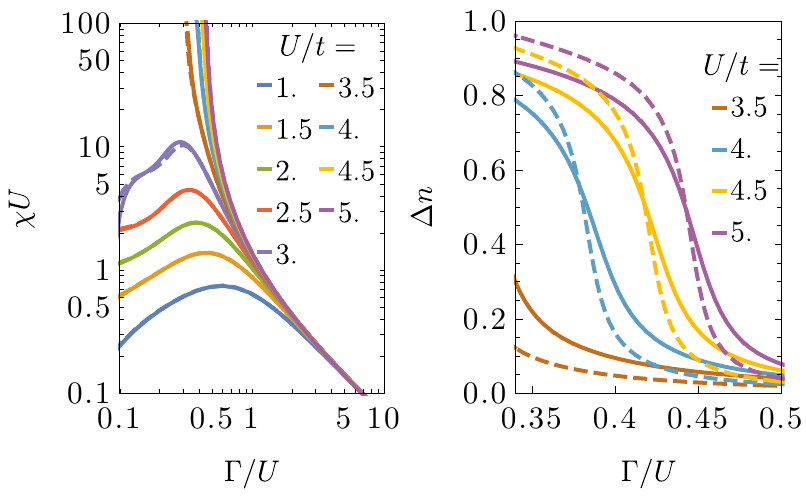}
        \put(0,60){(a)}
        \put(50,60){(b)}
    \end{overpic}
    \caption{FRG results for (a) the susceptibility and (b) the CDW order parameter as a function of the reservoir coupling $\Gamma$ at a fixed electrical field \(E/t=0.2\) which drives the system out of equilibrium for various values of $U$. Solid and dashed lines show data for $s/t=0.02$ and $s/t=0.01$, respectively. The FRG calculation was carried out using $M=4$. The field $E$ induces a current, which suppresses charge-ordering tendencies: The system is in a disordered phase (finite susceptibility) for any $\Gamma$ at $U/t\lesssim3$; for larger $U$, we find a transition into a CDW phase at a finite value of $\Gamma$.
    }\label{fig:finite_e_phase_trans}
\end{figure}

\begin{figure}
    \begin{overpic}[width=\columnwidth]{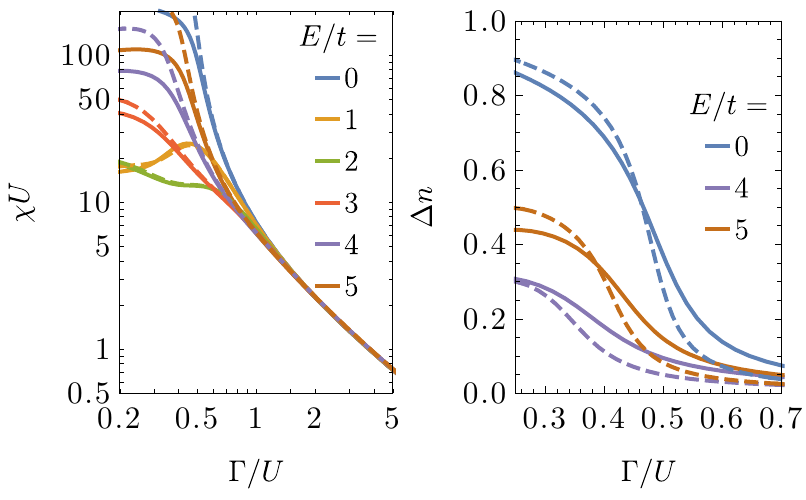}
        \put(0,60){(a)}
        \put(50,60){(b)}
    \end{overpic}
    \caption{The same as in Fig.~\ref{fig:finite_e_phase_trans} but for fixed \(U/t=5\) and varying electrical fields. Solid and dashed lines show data for $s/U=0.02$ and $s/U=0.01$, respectively. For \(E/t=1,2\), the system is in a disordered phase for any \(\Gamma\). For both smaller and larger electric fields, we find a transition into a CDW phase at a finite value of \(\Gamma/U\). The system shows reentrance behavior.
    }\label{fig:finite_u_phase_trans}
\end{figure}

We now turn to the phase diagram of the system driven out of equilibrium by a finite electric field $E>0$. Results for the susceptibility as well as for the CDW order parameter are shown in Fig.~\ref{fig:finite_e_phase_trans} and \ref{fig:finite_u_phase_trans} for a constant electric field $E/t=0.2$ and a constant interaction $U/t=5$, respectively. 

For a constant, small electric field of $E/t=0.2$, the system is in a disordered phase for $U/t\lesssim 3$ at any value of $\Gamma$. However, the susceptibility does not decrease monotonically with $\Gamma$ but features novel structures which reflect the emergence of the multiple energy scales away from equilibrium [see Fig.~\ref{fig:finite_e_phase_trans}(a)]. The critical interaction beyond which we observe a transition into a CDW phase is drastically enhanced (here by roughly a factor of 2) compared to the case of $E=0$; the electric field drives a current through the system, and the tendencies to form charge order are suppressed.

In Fig.~\ref{fig:finite_u_phase_trans}, we show results for constant $U/t=5$ and various $E$. Small fields $E/t\lesssim2$ induce a current and thus reduce the tendency to form charge order. However, we observe a re-entrance into the CDW phase as the strength of the field is increased: For large $E$, the current is suppressed due to localization and can no longer inhibit CDW order.

\section{Conclusion}

We have discussed a method to treat an infinite quantum system which is driven out of thermal equilibrium in a translation-invariant fashion. In particular, we have set up recursion relations for Keldysh-Schwinger Green's functions and demonstrated how they can be solved efficiently in a numerically-exact way. The algorithm is directly applicable to any diagrammatic Green's function based method such as (dynamical) mean field theories or the functional renormalization group. Furthermore, the presented iterative Green's functions scheme has direct relevance to periodically driven systems described by a Floquet Green's functions approach.\cite{Aoki2014,AK1,AK2}

We applied this general machinery within a novel, second-order Keldysh formulation of the functional renormalization group; significant effort was devoted to efficiently implementing the FRG flow equation via fast Fourier transforms. As a physically relevant example, we studied a tight-binding chain of interacting spinless fermions coupled to reservoirs and driven out of equilibrium by an electric field. It is known that the closed system features an equilibrium Berezinskii-Kosterlitz-Thouless phase transition into a CDW phase beyond a critical interaction strength $U_\tn{crit}$. In contrast to mean-field approaches, the FRG reproduces this result and can be used to determine $U_\tn{crit}$ in the presence of a finite reservoir coupling. A small electric field induces a current and thus suppresses tendencies to form charge order. For large field, however, Wannier-Stark localization leads inhibits currents and leads to a re-entrance into the CDW phase.

\section{Acknowledgments} DMK was funded by the Deutsche Forschungsgemeinschaft (DFG, German Research Foundation) under Germany's Excellence Strategy - Cluster of Excellence Matter and Light for Quantum Computing (ML4Q) EXC 2004/1 - 390534769. We acknowledge support from the Max Planck-New York City Center for Non-Equilibrium Quantum Phenomena. CKa and CKl acknowledge support by the Deutsche Forschungsgemeinschaft through the Emmy Noether program (KA 3360/2-1).

\appendix
\section{Solution of non-local self-consistency equations}\label{sec:sol_self_cons}
In sections~\ref{sec:iterative_gf} and~\ref{sec:iterative_s} we derived self-consistency equations for the auxiliary Green's functions and single-scale propagators [see Eqs.~\eqref{eq:iter_gf2}, \eqref{eq:iter_gk2}, \eqref{eq:iter_sr2}, and \eqref{eq:iter_sk2}] of the form:
\begin{equation}\label{eq:nonloc_self_cons}
    G(\omega\pm NE)=\mathcal{F}\left( G(\omega)\right),
\end{equation}
where \(G\in\left\{\calrLR_{\li\li},\calkLR_{\li\li},\calrLR_{\li\li},\calSrLR_{\li\li},\calSkLR_{\li\li} \right\}\).
At vanishing \(E=0\), such equations can be solved with a self-consistency loop. For \(E>0\), Eq.~(\ref{eq:nonloc_self_cons}) is non-local in the frequency, and we will now discuss to to solve such an equation efficiently. In all cases, the following boundary condition holds:
\begin{equation}\label{eq:nonloc_con_bound}
    \lim_{\omega\to \pm\infty}G(\omega)=0.
\end{equation}
Together, Eq.~\eqref{eq:nonloc_self_cons} and~\eqref{eq:nonloc_con_bound} uniquely define \(G(\omega)\).
\paragraph{Equidistant grids of spacing \(NE\)}
The first, straightforward way to solve such equation is to consider an equidistant grid
\begin{equation}
    \left\{-k_\mathrm{max}NE,(-k_\mathrm{max}+1)NE,\dots,k_\mathrm{max}NE \right\}.
\end{equation}
With the initial assumption \(G(\pm k_\mathrm{max}NE)=0\), Eq.~(\ref{eq:nonloc_self_cons}) can be used to successively calculate the Green's function on the entire grid.
\paragraph{Arbitrary grids}
The method described above becomes inefficient if \(NE\) is much smaller than the required grid spacing.
Alternatively, we can consider an arbitrary frequency discretization \(\{\omega_1,\dots,\omega_{k_\mathrm{max}}\}\) and assume \(G(\omega_{k_\mathrm{max}})=G(\omega_1)=0\). For simplicity, we restrict us to the case of negative sign in Eq.~(\ref{eq:nonloc_self_cons}). In that case one uses the following recursive algorithm to successively obtain $G(\omega)$ on the entire grid. One assumes that $G(\omega)$ has already been calculated for all frequencies $\omega_{k+1},\ldots,\omega_{k_\tn{max}}$. In order to compute $\omega_{k}$, we find the largest frequency $\omega_i$ which fulfills
\begin{equation}
 \omega_i\leq\omega_{k}+NE.
\end{equation}
Note that one always has $i\geq k$. If we assume that \(G\) is well approximated by a linear interpolation between grid points, we obtain:
\begin{equation}\label{eq:nonloc_iter1}
    \begin{split}
                                    G(\omega_k) & = \mathcal{F}\left[G(\omega_k+NE)\right]\\ & \approx
                                    \mathcal{F}\left[(1-\delta)G(\omega_i)+\delta G(\omega_{i+1})\right],
    \end{split}
\end{equation}
where
\begin{equation}
    \begin{split}
                                    &\delta=\frac{\omega_k+NE-\omega_i}{\omega_{i+1}-\omega_i}.\\
    \end{split}
\end{equation}
Whenever \(k=i\), Eq.~(\ref{eq:nonloc_iter1}) is solved using a self-consistency loop, otherwise \(G(\omega_i)\) has previously been computed [\(G(\omega_{i+1})\) is known since $i\geq k$]. If $i=k_\tn{max}$, we set $G(\omega_{i+1})=0$.

\section{Efficient convolution}\label{sec:convo}
For a numerical implementation of the FRG algorithm discussed in this work, it is essential to perform the integrals appearing on the rhs of the flow equations efficiently. As mentioned above, this can be achieved by treating all integrals as convolutions and by utilizing an efficient algorithm to perform these.
To that end, we discretize all frequencies on a grid as discussed in Sec.~\ref{sec:freq_disc}. In the following, $N$ denotes the total number of frequency points.

We define the convolution of two functions as
\begin{equation}
 (f*g)(y)=\int_{-\infty}^\infty \dOp x f(x)g(y-x).
\end{equation}
If this integral is carried out naively, this requires \(\ord{N}\) operations, and the effort to obtain the rhs of all of the flow equations thus scales as \(\ord{N^2}\). In this Appendix, we will discuss how such convolutions can be obtained efficiently on equidistant grids, on grids with an arbitrary spacing, and on mixed grids (which are required within the FRG).

\subsection{Equidistant grids}
\label{sec:equiGrid}
For equidistant grids, we can rewrite the convolution in terms of a discrete Fourier transform, which will allow us to perform significantly faster computations.
\subsubsection{Discretization}
\begin{figure}
    \begin{center}
        \begin{overpic}[width=1.0\columnwidth]{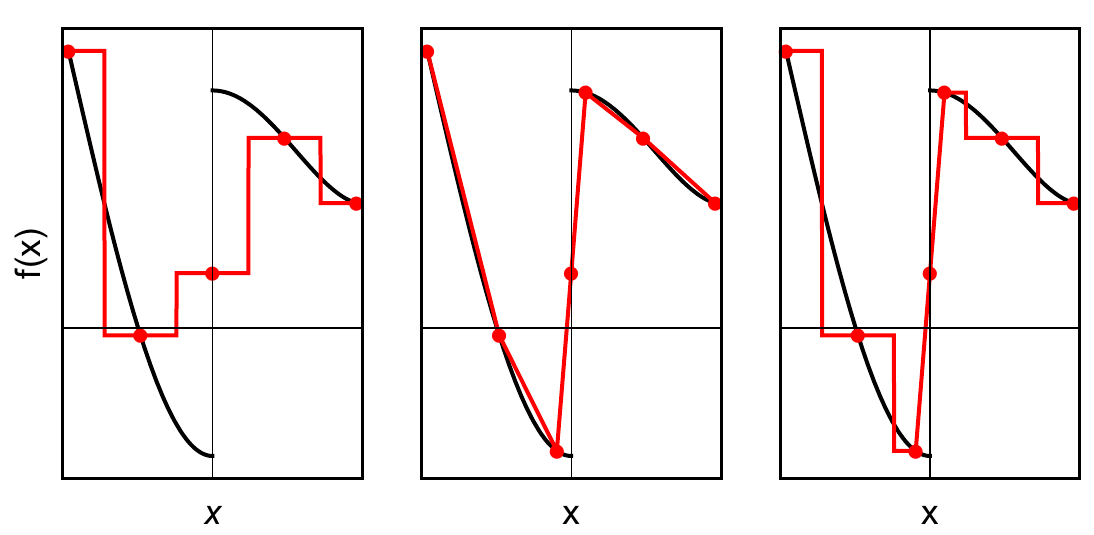}
        \end{overpic}
    \end{center}
\caption{
    Pictorial representation of the three different discretizations discussed in this Appendix. From left to right we show a piecewise constant approximation on an equidistant grid (discussed in~\ref{sec:equiGrid}), a piecewise linear approximation on an arbitrary grid (see~\ref{sec:arb_grid}) and a piecewise constant approximation with piecewise linear refinements (see~\ref{sec:mixed_grids}).
}
\label{fig:discret_convo}
\end{figure}

We employ an equidistant grid with \(N\in 2\mathbb{N}+1\) points and approximate \(f\) and \(g\) by piecewise constant functions (compare left panel of Fig.~\ref{fig:discret_convo}):
\begin{equation}
    \label{eq:pcons}
\tilde h(x)=\left\{
    \begin{matrix}
    0  &                             &  &x&<&-\frac{N}{2}w\\
    h_i&\left(i-\frac{1}{2}\right)w &\leq&x&<&\left(i+\frac{1}{2}\right)w\\
    0  &\frac{N}{2}w &\leq&x& &
    \end{matrix}
    \right.
\end{equation}
with a vector \(\left(h_i\right)_{i=\left\{-\frac{N-1}{2}\dots\frac{N-1}{2}\right\}}\), and \(h_i=h\left(iw\right)\) and \(h=f,g\).
A convolution of \(f,g\) then reduces to a convolution of vectors:
\begin{equation}
(f*g)(iw)=\int_{-\infty}^\infty \dOp x f(x)g(y-x)\approx w\sum_k f_k g_{i-k},
\end{equation}
with the summation bounded appropriately. The summation runs over \(\ord{N}\) elements, and it thus takes \(\ord{N^2}\) operations to compute the convolution on the entire original grid (which has \(N\) points). This can be improved by employing a Fourier transform.

\subsubsection{Fourier transform}
Using the discrete Fourier transform
\begin{equation}
    \bar f_k=\frac{1}{\sqrt{N}}\sum_l e^{\I lk\frac{2\pi}{N}}f_l, \hspace{0.2cm}
f_l=\frac{1}{\sqrt{N}}\sum_k e^{-\I lk\frac{2\pi}{N}}\bar f_k,
\end{equation}
one rewrites the discrete convolution as
\begin{equation}
\begin{split}
    w\sum_j f_jg_{n-j}&=\frac{w}{N}\sum_{k_1,k_2}\sum_j e^{-\I j k_1\frac{2\pi}{N}} \bar f_{k_1} e^{-\I(n-j)k_2\frac{2\pi}{N}} \bar g_{k_2}\\
                     &=w\sum_{k}e^{-\I n k\frac{2\pi}{N}} \bar f_{k} \bar g_{k},
\end{split}
\end{equation}
which is the \(n\)-th element of the back-transform of the product of \(\bar f_k \bar g_k\).

By employing the \emph{fast Fourier transform} (FFT) algorithm, the discrete Fourier transform can be computed in \(\ord{N\log(N)}\) operations,\footnote{In our implementation of the algorithm we employ the FFTW library,\cite{Frigo1999} which provides a FFT implementation for discrete Fourier transforms. To obtain optimal performance, it is beneficial to chose a grid size with small prime factors; as our grid contains an odd number of points we chose a power of \(3\).} which allows us to obtain \emph{all} components of the convolution on the same grid in \(\ord{N\log(N)}\) operations.

While this algorithm is very efficient, it is specifically designed for equidistant grids and can not easily be generalized to more general discretizations. For our application with its vastly different energy scales, one is forced to use an overly dense grid, which diminishes the advantage of this method.

\subsection{Arbitrary grids}
\label{sec:arb_grid}
Alternatively, one can work with an entirely arbitrary grid defined via  \(\left\{ x_1, \dots, x_N\right\}\) with \(x_1<\dots<x_N\). We assume that the functions $f,g$ are approximated piecewise linearly (compare center panel of~Fig.~\ref{fig:discret_convo}),
\begin{equation}
    \label{eq:plin}
\tilde h(x)=\left\{
    \begin{matrix}
    0  &                             &  &x&<&x_1\\
    a^h_ix+b^h_i &x_i&\leq&x&<&x_{i+1}\\
    0  &x_N&\leq&x,& &
    \end{matrix}
    \right.
\end{equation}
 with \(a_i x_{i+1} +b_i= a_{i+1} x_{i+1}+ b_{i+1}\ \forall i\), and
 \begin{equation}
     \tilde h(x_i)=h(x_i)\ \forall i\in \{1,\dots,N\}
 \end{equation} 
 for \(h=f,g\). The convolution of the two functions can then be written as
\begin{equation}\label{eq:convo_arb}
    \begin{split}
        (f*g)(x_i)=&\int_{-\infty}^\infty \dOp x f(x)g(x_i-x)\\
        =\sum_{k,k'=1}^N& \int_{S_{kk'}} \dOp x \left(a^f_kx-b^f_k\right)\left(a^g_{k'} (x_i-x) -b^g_{k'}\right),\\
        S^i_{kk'}&={[x_k,x_{k+1}]\cap [x_i-x_{k'+1},x_i-x_{k'}]}.
    \end{split}
\end{equation}
Depending on the chosen discretization, the support \(S_{kk'}\) of the integrand at a given \(k\) might be non-zero for more than one \(k'\); however,
\begin{equation}
    \left|\left\{(k,k')\in \{1\dots N\}^2 \middle| S^i_{kk'}\neq \emptyset \right\}\right|\in \mathcal{O}(N).
\end{equation}
Given a fixed \(k\), the values of \(k'\) contributing to the rhs of Eq.~(\ref{eq:convo_arb}) can be identified in \(\ord{\log(N)}\) operations by traversing the \(x_i\) with an appropriate algorithm.
Using the indefinite integral
\begin{equation}
    \begin{split}
    &\int \dOp x\ (ax-b)(a'x-b')\\
    &=\frac{1}{3} aa'x^3+\frac{1}{2}\left(ab'+ba'\right) x^2+bb'x + c,
\end{split}
\end{equation}
the remaining convolutions of linear functions can be easily evaluated.

Note that for an efficient algorithm it is of crucial importance that the grid of \(x_i\) is sorted and that appropriate algorithms to identify the contributing \(k'\) are employed. On an arbitrary grid with \(N\) points, this algorithm requires \(\ord{N^2\log N}\) operations to obtain the convolutions on the entire grid.

While this algorithm allows for grids that take the vastly different energy scales of the problem into account, it is (compared with the case of equidistant grids) slow and is the bottleneck of such an implementation.

\subsection{Mixed grids}
\label{sec:mixed_grids}

\begin{figure*}
    \[
        \raisebox{-0.5\height}{\includegraphics[scale=0.6]{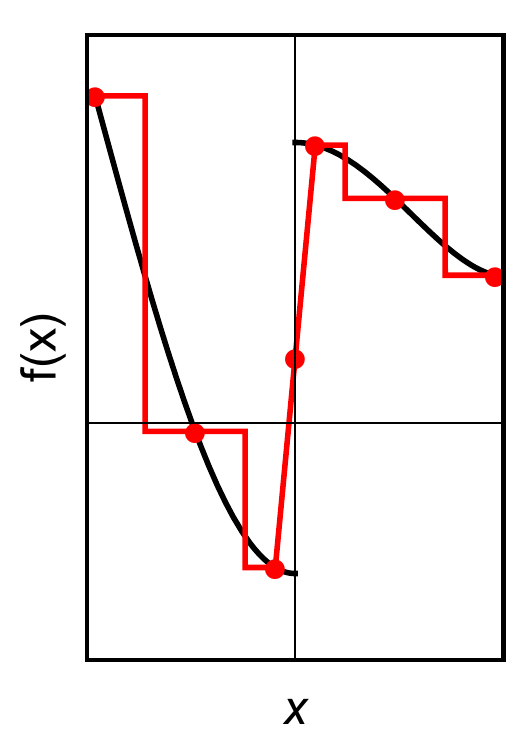}}=
        \raisebox{-0.5\height}{\includegraphics[scale=0.6]{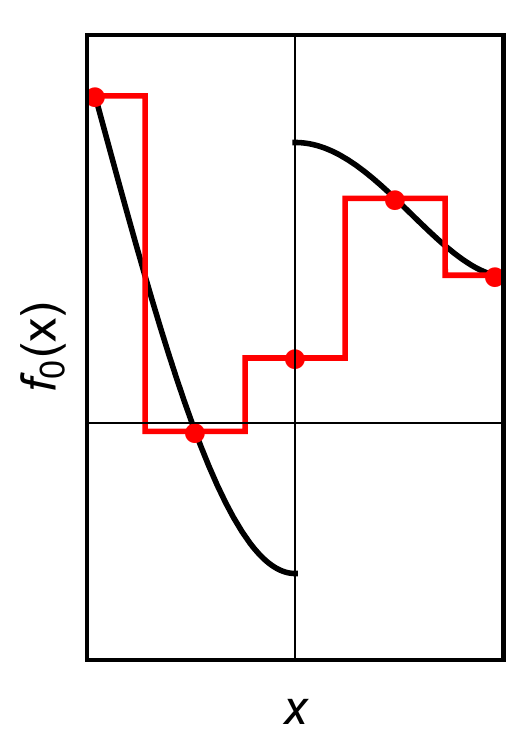}}+
        \raisebox{-0.5\height}{\includegraphics[scale=0.6]{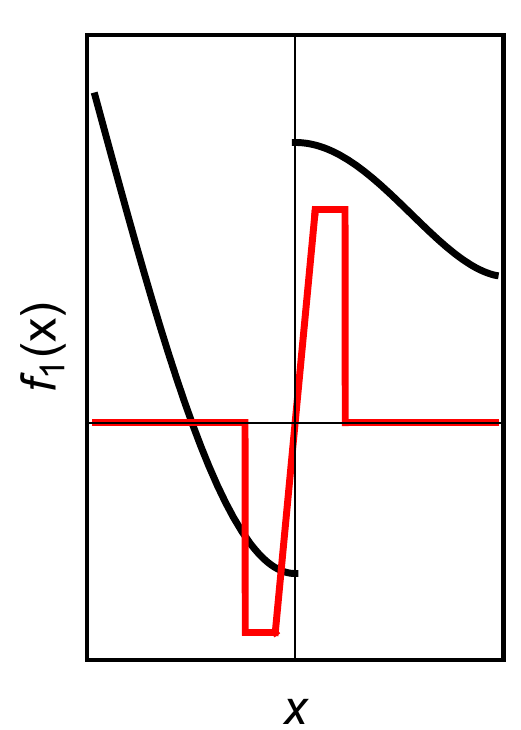}}
    \]
\caption{
    Pictorial representation of the decomposition proposed in Sec.~\ref{sec:arb_grid}. We obtain a approximation on a refined equidistant grid as a decomposition into a piecewise constant function \(f_0(x)\) and a series of piecewise linear functions (here only \(f^1_\mathrm{lin}\)) with support only within one segment of the equidistant grid. We have the freedom to choose \(\int \dOp x f^1_\mathrm{lin}(x)=0\).
}
\label{fig:discret_convo_decomp}
\end{figure*}

Let us assume that the functions \(f\), \(g\) are known on a grid \(\Omega\) containing \(\mathcal{O}(N)\) equidistant points as well as \(\mathcal{O}(k)\) additional points at arbitrary positions,
\begin{equation}
    \Omega=\left\{-\frac{N-1}{2}w, \dots, \frac{N-1}{2}w\right\}\cup \left\{x_1,\dots, x_k \right\},
\end{equation}
with \(-\frac{N-1}{2}w<x_1<x_2<\cdots<x_k<\frac{N-1}{2}w\). We first analyze sections of the equidistant subgrid.

\subsubsection{Within one section}
For a given \(n\in \left\{-\frac{N-1}{2},\dots,\frac{N-1}{2}\right\}\) consider the adjacent points 
\begin{equation}
S_n=\left(nw-\frac{w}{2}, nw+\frac{w}{2}\right)\cap \left\{x_1,\dots,x_k\right\}. 
\end{equation}
Then there is exactly one decomposition \(f(x)\approx a^n+f^n_\mathrm{lin}(x)\) fulfilling the following three conditions: Firstly, we require
\begin{equation}
    f(x)=a^n+f^n_\mathrm{lin}(x)\ \forall x\in\{nw\}\cup S_n,
\end{equation}
where \(f_\mathrm{lin}\) is a piecewise linear function with support only in \(\left(nw-\frac{w}{2}, nw+\frac{w}{2}\right)\). Secondly, we require 
\begin{equation}
    \frac{\dOp f^n_\mathrm{lin}(x)}{\dOp x}\to 0 \text{ for } x\to nw\pm \frac{w}{2}.
\end{equation}
Lastly, we impose the condition
\begin{equation}
    \int_{-\infty}^\infty \dOp x f^n_\mathrm{lin}(x)=\int_{nw-\frac{w}{2}}^{nw+\frac{w}{2}} \dOp x f^n_\mathrm{lin}(x)=0.
\end{equation}
If this was not fulfilled, the finite weight of \(f_\mathrm{lin}\) could be absorbed into the constant \(a^n\).

\subsubsection{On the entire grid}
Following this procedure in every segment results in a representation
\begin{equation}
    \label{eq:decomp_convo}
    f(x)\approx f_0(x)+\sum_{n=1}^{N} f^n_\mathrm{lin}(x)
\end{equation}
with
\begin{equation}
f_0(x)=\left\{
    \begin{matrix}
    a_n&|x-nw|<\frac{w}{2}\\
    0  &\text{otherwise}.
    \end{matrix}
    \right.
\end{equation}
Note that only a small number of \(f^n_\mathrm{lin}(x)\not\equiv0\) (in fact, \(k\) at most), since these linear functions are only required if one of the additional points \(x_i\) falls into the interval \(\left(nw-\frac{w}{2}, nw+\frac{w}{2}\right)\).

A visual example of such a decomposition is shown in Fig.~\ref{fig:discret_convo_decomp}, a concrete example is presented in the next section.
\subsubsection{Example}
As an example, consider
\begin{equation}
    f(x)=\mathrm{sgn}(x-0.2)
\end{equation}
on the grid
\begin{equation}
    \Omega=\{-1,0,1\}\cup \{0.1,0.3\}.
\end{equation}
These functions can be approximated as
\begin{equation}
    \begin{split}
    f(x)&\approx f_0(x)+f^1_\mathrm{lin}(x),\\[1ex]
    \end{split}
\end{equation}
with
\begin{equation}
    \begin{split}
    f_0(x)&=\left\{\begin{matrix}
        -1 & -1.5\leq x < -0.5\\ 
    -0.4 & -0.5\leq x <0.5\\
    1. & 0.5\leq x <1.5\\
0 & \text{otherwise}\end{matrix}  \right.
    \end{split}
\end{equation}
and
\begin{equation}
    \begin{split}
f^1_\mathrm{lin}(x)&=\left\{\begin{matrix}
        -0.6 & -0.5\leq x <0.1\\
        10x-1.6 & 0.1\leq x <0.3\\
        1.4 & 0.3\leq x <0.5\\
0 & \text{otherwise.}\end{matrix}  \right. 
    \end{split}
\end{equation}

\subsubsection{Performing convolutions on such approximations}
Using its linearity, a convolution of two functions approximated in such a way can always be decomposed as:
\begin{equation}
    \begin{split}
        (f*g)(y)&=(f_0*g_0)(y)+\sum_i(f^i_\mathrm{lin}*g_0)(y)\\
                &\hspace{1cm}+\sum_j(f_0*g_\mathrm{lin}^j)(y)+\sum_{i,j}(f_\mathrm{lin}^i*g^j_\mathrm{lin})(y)\\
    \end{split} 
\end{equation}
We now evaluate this decomposition on the grid and have to distinguish between two cases:

\paragraph{For all \(y\) in the equidistant grid}
the first term can be computed as outlined in Sec.~\ref{sec:equiGrid} in \(\ord{N\log(N)}\) operations.
Due to our choice \(\int \dOp x f^i_\mathrm{lin}(x)=0=\int \dOp x g^i_\mathrm{lin}(x)\), the support of these functions as well as the fact that \(f_0\) and \(g_0\) are piecewise constant, the second and third term vanish.
The last term can be explicitly computed in \(\ord{k^2\log(N)}\) operations: the summands are only non-zero for \(\ord{k^2}\) values of \((i,j)\), and for each such combination \(\ord{\log(N)}\) operations are required to identify the (small) support of \(f_\mathrm{lin}^i*g^j_\mathrm{lin}\).
\paragraph{For each \(y\) outside the equidistant grid}
one can obtain the convolution explicitly using the algorithm for arbitrary grids discussed in Sec.~\ref{sec:arb_grid}; as there are only \(k\) such points, this results in an algorithm of \(\ord{kN\log(N)}\) operations.

\bibliography{ref}

\end{document}